\documentclass[11pt]{article}
\usepackage{amsmath,amssymb,graphicx}
\usepackage{hyperref}

\newcommand{\eprint}[1]{{\href{http://arxiv.org/abs/#1}{\texttt{[#1}]}}}
\newcommand{\eprintN}[1]{{\href{http://arxiv.org/abs/#1}{\texttt{#1 [hep-th]}}}}
\newcommand{\eprintNM}[1]{{\href{http://arxiv.org/abs/#1}{\texttt{#1 [math.NT]}}}}

\usepackage{multicol,color,longtable}

\definecolor{darkred}{rgb}{0.65,0.15,0}
\hypersetup{pdfborder={0 0 0},colorlinks=true,urlcolor=darkred,citecolor=blue,linkcolor=darkred,linktocpage=true}

\usepackage{tikz}
\usetikzlibrary{arrows}
\usepackage{pgfplots}

\usepackage{mathrsfs}
\usepackage[Symbol]{upgreek}
\usepackage{dsfont}
\usepackage[vcentermath]{youngtab}
\usepackage{textcomp}

\def\cD{{\cal D}}
\def\cF{{\cal F}}
\def\cE{{\cal E}}
\def\cV{{\cal V}}

\def\cP{{\cal P}}

\newcommand{\WSOXVI}[8]{\medmuskip=-1mu\mbox{\tiny $ {   \biggl[ \, \begin{array}{cc} &\hspace{-2.9mm}#7 \vspace{ -1.5mm} \\  \hspace{-2mm}#1\hspace{0.9mm} #2\hspace{0.9mm}  #3 \hspace{0.9mm} #4 \hspace{0.9mm} #5\hspace{0.9mm} #6& \hspace{-0.9mm}\vspace{-1.5mm}\\ &\hspace{-2.9mm} #8  \end{array}\hspace{-1.5mm}\, \biggr] }$}\medmuskip=4mu plus 2mu minus 4mu}

\def\ie{{\it i.e.}\ }
\def\eg{{\it e.g.}\ }
\newcommand{\Scal}[1]{\Bigl ({#1} \Bigr )}
\newcommand{\scal}[1]{\bigl ({#1} \bigr )}
\newcommand{\nn}{\nonumber}
\def\be{\begin{equation}}\def\ee{\end{equation}}
\def\bea{\begin{eqnarray}}\def\eea{\end{eqnarray}}

\newcommand{\ints}{\mathds{Z}}
\newcommand{\reals}{\mathds{R}}
\newcommand{\cx}{\mathds{C}}

\newcommand{\CR}{\nonumber \\*}
\newcommand{\sfrac}[2]{{\scriptstyle \frac{#1}{#2}}}
\newcommand{\stfrac}[2]{{\scriptscriptstyle \frac{#1}{#2}}}

\newcommand{\tr}{\mathrm{tr}}

\newcommand{\mf}[1]{{\mathfrak{#1}}}

\newcommand{\cosel}{{\mathcal{V}}}
\newcommand{\gs}{g}

\newcommand{\ba}{/ \hspace{-1.2ex}} 
\newcommand{\baa}{/ \hspace{-1.4ex}}

\newcommand{\DEVIII}[8]{\medmuskip=-1mu\mbox{\tiny $ {   \biggl[ \, \begin{array}{ll} &\hspace{-2.5mm}#2 \vspace{ -0.5mm} \\  \hspace{-2mm}#1\hspace{0.9mm} #3&\hspace{-2.5mm}  #4 \hspace{0.9mm}  #5 \hspace{0.9mm}  #6 \hspace{0.9mm}  #7 \hspace{0.9mm}  #8  \end{array}\hspace{-2.0mm}\, \biggr] }$}\medmuskip=4mu plus 2mu minus 4mu}

\newcommand{\DEVII}[7]{\medmuskip=-1mu\mbox{\tiny $ {   \biggl[\,  \begin{array}{ll} &\hspace{-2.5mm}#2 \vspace{- 0.5mm}\\  \hspace{-2mm}#1\hspace{0.9mm} #3&\hspace{-2.5mm}  #4 \hspace{0.9mm}  #5 \hspace{0.9mm}  #6 \hspace{0.9mm}  #7  \end{array}\hspace{-2.0mm}\, \biggr] }$}\medmuskip=4mu plus 2mu minus 4mu}

\newcommand{\DEVI}[6]{\medmuskip=-1mu \mbox{\tiny $ {   \biggl[\,  \begin{array}{ccccc} &&\hspace{-3.9mm}#2 \vspace{0.5mm}&&\hspace{-3.9mm} \vspace{ -0.5mm} \\  \hspace{-2mm}#1&\hspace{-3.9mm} #3&\hspace{-3.9mm}  #4& \hspace{-3.9mm}  #5& \hspace{-3.9mm}  #6  \end{array}\hspace{-2.0mm}\, \biggr] }$}\medmuskip=4mu plus 2mu minus 4mu}

\newcommand{\DSOVI}[6]{\medmuskip=-1mu\mbox{\tiny $ {   \biggl[ \, \begin{array}{cc} &\hspace{-2.9mm}#5 \vspace{ -1.5mm} \\  \hspace{-2mm}#1\hspace{0.9mm} #2\hspace{0.9mm}  #3 \hspace{0.9mm} #4 & \hspace{-0.9mm}\vspace{-1.5mm}\\ &\hspace{-2.9mm} #6  \end{array}\hspace{-1.5mm}\, \biggr] }$}\medmuskip=4mu plus 2mu minus 4mu}

\newcommand{\DSOV}[5]{\medmuskip=-1mu\mbox{\tiny $ {   \biggl[ \, \begin{array}{cc} &\hspace{-2.9mm}#4 \vspace{ -1.5mm} \\  \hspace{-2mm}#1\hspace{0.9mm} #2\hspace{0.9mm}  #3& \hspace{-0.9mm}\vspace{-1.5mm}\\ &\hspace{-2.9mm} #5  \end{array}\hspace{-1.5mm}\, \biggr] }$}\medmuskip=4mu plus 2mu minus 4mu}

\newcommand{\DSOIV}[4]{\medmuskip=-1mu\mbox{\tiny $ {   \biggl[ \, \begin{array}{cc} &\hspace{-2.9mm}#3 \vspace{ -1.5mm} \\  \hspace{-2mm}#1\hspace{0.9mm} #2& \hspace{-0.9mm}\vspace{-1.5mm}\\ &\hspace{-2.9mm} #4  \end{array}\hspace{-1.5mm}\, \biggr] }$}\medmuskip=4mu plus 2mu minus 4mu}

\newcommand{\DSOIVS}[4]{\medmuskip=-1mu\mbox{\tiny $ {   \biggl[ \, \begin{array}{cc} &\hspace{-2.9mm}#1 \vspace{ -1.5mm} \\  \hspace{-2mm}#4\hspace{0.9mm} #2& \hspace{-0.9mm}\vspace{-1.5mm}\\ &\hspace{-2.9mm} #3  \end{array}\hspace{-1.5mm}\, \biggr] }$}\medmuskip=4mu plus 2mu minus 4mu}

\newcommand{\DSLIII}[3]{\medmuskip=-1mu\mbox{\tiny $ {   \left[ \hspace{0mm}#1\hspace{0.9mm} #2\hspace{0.9mm}  #3 \hspace{0mm}\right] }$}\medmuskip=4mu plus 2mu minus 4mu}

\newcommand{\DSLII}[2]{\medmuskip=-1mu\mbox{\tiny $ {   \left[ \hspace{0mm}#1\hspace{0.9mm} #2 \hspace{0mm}\right] }$}\medmuskip=4mu plus 2mu minus 4mu}

\newcommand{\DSLI}[1]{\medmuskip=-1mu\mbox{\tiny $ {   \left[ \hspace{0mm}#1 \hspace{0mm}\right] }$}\medmuskip=4mu plus 2mu minus 4mu}

\newcommand{\gra}[2]{{\scriptscriptstyle (#1 , #2 )}}
\newcommand{\ord}[1]{{\scriptscriptstyle (#1)}}

\makeatletter

\@addtoreset{equation}{section}
\makeatother

\begin{document}

\thispagestyle{empty}

{\flushright {CPHT-RR017.0515}\\[15mm]}

\begin{center}
{\LARGE \bf Supergravity divergences, supersymmetry\\[5mm] and automorphic forms}\\[10mm]

\vspace{8mm}
\normalsize
{\large  Guillaume Bossard${}^{1}$ and Axel Kleinschmidt${}^{2,3}$}

\vspace{10mm}
${}^1${\it Centre de Physique Th\'eorique, Ecole Polytechnique, CNRS\\
91128 Palaiseau cedex, France}
\vskip 1 em
${}^2${\it Max-Planck-Institut f\"{u}r Gravitationsphysik (Albert-Einstein-Institut)\\
Am M\"{u}hlenberg 1, DE-14476 Potsdam, Germany}
\vskip 1 em
${}^3${\it International Solvay Institutes\\
ULB-Campus Plaine CP231, BE-1050 Brussels, Belgium}

\vspace{20mm}

\hrule

\vspace{10mm}

\begin{tabular}{p{12cm}}
{\small
We investigate the occurrence of divergences in maximal supergravity in various dimensions from the point of view of supersymmetry constraints on the U-duality invariant threshold functions defining the higher derivative couplings in the effective action. Our method makes use of tensorial differential equations that constrain these couplings. We study in detail the Fourier modes and wave-front sets of the associated automorphic functions and find that they are always associated to special nilpotent orbits.

}
\end{tabular}
\vspace{7mm}
\hrule
\end{center}

\newpage
\setcounter{page}{1}

\tableofcontents

\vspace{5mm}
\hrule
\vspace{5mm}

\section{Introduction}

The four-graviton scattering amplitude in type II string theory in $D$ dimensions has an analytic part that possesses a low-energy expansion of the form
\begin{align}
\mathcal{A}^{\textrm{analytic}}_D (s,t,u;\cosel_D) = \left( 3 \sigma_3^{-1} + \sum_{p,q=0}^\infty \mathcal{E}_{\gra{p}{q}}^{\ord{D}}(\cosel_D) \sigma_2^p \sigma_3^q\right) \ell_D^6 \mathcal{R}^4,
\end{align}
where $\ell_D$ is the $D$-dimensional Planck length, $s,t,u$ are the standard Mandelstam variables, and $\sigma_k =\left( \ell_D^2/4\right)^k(s^k+t^k+u^k)$ dimensionless combinations of them. $\mathcal{R}^4$ denotes a specific contraction of the four polarisations and momenta of the four graviton states using the so-called $t_8t_8$ tensor~\cite{Green:1981ya,Gross:1986iv}. The most important objects in the above equation are the functions $\mathcal{E}_{\gra{p}{q}}^{\ord{D}}(\cosel_D)$ that depend on the moduli $\cosel_D\in E_{11-D}/K(E_{11-D})$, where $E_{11-D}$ is the Cremmer--Julia symmetry group in $D$ dimensions~\cite{Cremmer:1979up} and $K(E_{11-D})$ its maximal compact subgroup. The functions $\mathcal{E}_{\gra{p}{q}}^{\ord{D}}(\cosel_D)$ must be invariant under the discrete U-duality group $E_{11-D}(\ints)$~\cite{Hull:1994ys,Green:1997tv}. Moreover, the functions have to satisfy differential constraints from supersymmetry~\cite{Pioline:1998mn,Green:1998by,Green:2005ba,Green:2010wi,Bossard:2014lra,Bossard:2014aea,Bossard:2015uga,Wang:2015jna,Wang:2015aua} that can also be understood representation-theoretically~\cite{Green:2010kv,Pioline:2010kb,Green:2011vz}. The differential equations include Poisson-type equations of the form~\cite{Green:2010wi,Pioline:2015yea}
\begin{subequations}
\label{Epq}
\begin{align}
\label{R4eqn}
\left( \Delta - \frac{3(11-D)(D-8)}{D-2}\right) \mathcal{E}_{\gra00}^{\ord{D}}  &= 6\pi \delta_{D,8},\\
\label{D4R4eqn}
\left( \Delta - \frac{5(12-D)(D-7)}{D-2}\right) \mathcal{E}_{\gra10}^{\ord{D}}  &= 40\zeta(2) \delta_{D,7}+7\mathcal{E}_{\gra00}^{\ord{4}} \delta_{D,4},\\
\label{D6R4eqn}
\left( \Delta - \frac{6(14-D)(D-6)}{D-2}\right) \mathcal{E}_{\gra01}^{\ord{D}}  &=-\left(\mathcal{E}_{\gra00}^{\ord{D}}\right)^2+ 40\zeta(3) \delta_{D,6}+\frac{55}3\mathcal{E}_{\gra00}^{\ord{5}} \delta_{D,5} + \frac{85}{2\pi} \mathcal{E}_{\gra10}^{\ord{4}} \delta_{D,4}
\end{align}
\end{subequations}
for the first three functions $\mathcal{E}_{\gra00}^{\ord{D}}$, $\mathcal{E}_{\gra10}^{\ord{D}}$ and $\mathcal{E}_{\gra01}^{\ord{D}}$ that are associated with $R^4$, $\nabla^4R^4$ and $\nabla^6R^4$ type corrections to the supergravity action, respectively, and are of $\tfrac12$-, $\frac14$- and $\tfrac18$-BPS-type, respectively.

Our interest here lies in the dimension-dependent source terms in~\eqref{D6R4eqn} for the $\nabla^6R^4$ coefficient $\mathcal{E}_{\gra01}^{\ord{D}}$ that we will determine by a new method in this paper. The source terms are related to perturbative divergences in supergravity. This is most clearly visible in the string perturbation expansion of the function $\mathcal{E}_{\gra01}^{\ord{D}}$ itself. The result given in~\cite{Pioline:2015yea} for the non-analytic part is\footnote{We have denoted the $D$-dimensional string coupling by $\gs_D$. When it is clear from the context which dimension we are in, we will often omit the subscript.} 
\begin{align}
\label{E01pert}
\mathcal{E}_{\gra01}^{\ord{D}}  &= \left(\frac{4\pi^2}{27}  \log^2 \gs_8  + \frac{2\pi}9\left(\frac{\pi}2+\mathcal{E}_{\gra00}^{(8),\textrm{an}}\right)\log \gs_8 \right) \delta_{D,8}\nn\\
&\quad + 5\zeta(3) \log \gs_6 \delta_{D,6} + \frac{20}{9} \mathcal{E}_{\gra00}^{\ord{5}} \log \gs_5 \delta_{D,5} + \frac{5}{\pi}\mathcal{E}_{\gra10}^{\ord{4}}  \log\gs_4 \delta_{D,4}+\ldots.
\end{align}
The term in $\delta_{D,6}$ is related to a three-loop ultraviolet logarithmic divergence of the supergravity four-graviton scattering~\cite{Green:2010sp} and its value agrees with the field theory result of~\cite{Bern:2008pv} as shown in~\cite{DHoker:2014gfa,Pioline:2015yea}. The terms in $\delta_{D,5}$ and $\delta_{D,4}$ are related to form factor divergences in supergravity. The precise coefficients in~\eqref{D6R4eqn} have only appeared recently in~\cite{Pioline:2015yea} and we will present here an independent derivation of these coefficients based on the tensorial differential equations of~\cite{Bossard:2014lra,Bossard:2014aea,Bossard:2015uga}.

There is a deep connection between constraints from supersymmetry and the Fourier modes of the functions $\mathcal{E}_{\gra{p}{q}}^{\ord{D}}$~\cite{Pioline:2010kb,Green:2011vz}. The Fourier modes arise in perturbative expansions of the type~\eqref{E01pert} where one exploits the periodicity of  $\mathcal{E}_{\gra{p}{q}}^{\ord{D}}$ under discrete (Peccei--Quinn type) shift symmetries of some axionic moduli. The non-zero Fourier modes in such expansions contain the non-perturbative corrections to the scattering process as they arise from instantons in string theory~\cite{Green:1997tv}.  For the BPS-type correction terms in~\eqref{Epq}, only specific supersymmetric instantons can contribute and this puts restrictions on the structure of the Fourier expansion. Mathematically, this is reflected in the so-called {\em wave-front set} of the $E_{11-D}(\ints)$ invariant functions $\mathcal{E}_{\gra{p}{q}}^{\ord{D}}$. The wave-front set is a union of nilpotent orbits of the group $E_{11-D}$ acting on its Lie algebra (see for example~\cite{Moeglin,Ginzburg,Jiang:2014}). As Fourier modes can be associated with nilpotent orbits, the structure of the wave-front set captures the structure of the Fourier expansion and the correction terms $R^4$, $\nabla^4 R^4$ and $\nabla^6 R^4$ can be associated with points on the Hasse diagram of nilpotent orbits~\cite{Green:2011vz,Bossard:2014lra}. An important point that we will bring out in our discussion is that only so-called {\em special} nilpotent orbits are of relevance~\cite{Green:2011vz,Jiang:2014}. This will be discussed in detail for the case of $SO(5,5)$ which is the Cremmer--Julia group in $D=6$ dimensions. We analyse carefully the Fourier expansion of certain Eisenstein series on this group that arise in the derivation of the logarithmic divergences in~\eqref{Epq} and~\eqref{E01pert}, presenting among other things the Fourier modes of the spinor Eisenstein series. 

This article is structured as follows. In section~\ref{sec:Eisenstein} we review Eisenstein series on symmetric space $G/K$ as these are our main tools for constructing the correction terms  $\mathcal{E}_{\gra{p}{q}}^{\ord{D}}$. We also introduce tensorial differential operators that are needed for writing the supersymmetry constraints on the correction functions and Fourier modes in the subsequent chapters. In section~\ref{sec:SUSYE01}, we present and solve the supersymmetry constraints in the case of  $\mathcal{E}_{\gra{0}{1}}^{\ord{6}}$ that is related to the three-loop divergence in $D=6$ supergravity. Section~\ref{sec:Divs} contains a new method for finding the divergent terms in other dimensions and derives the coefficients in~\eqref{D6R4eqn} and~\eqref{E01pert} from a particular `adjoint' Eisenstein series on $E_{8(8)}$. Section~\ref{sec:Fourier} then analyses in detail the Fourier expansions of various Eisenstein series on $SO(5,5)$ in connection to the supersymmetric corrections. Two supplementary appendices contain details on the adjoint $E_{8(8)}$ series and on the Fourier expansion of the spinor series of $SO(5,5)$.

\section{Eisenstein series and tensorial differential operators}
\label{sec:Eisenstein}

In this paper, functions on the moduli space of string theory in $D$ space-time dimensions play a central role. The moduli space is a symmetric space $G/K$ with $G=E_{11-D}$ the Cremmer--Julia symmetry group in $D$-dimensional ungauged maximal supergravity and $K$ its maximal compact subgroup. On this space, we will define Eisenstein series invariant under U-duality $E_{11-D}(\ints)$ and tensorial differential operators that help to express the constraints from supersymmetry on functions on this space.

\subsection{Brief reminder of Eisenstein series}

We will use the following convention for Eisenstein series on a split real Lie group $G$, \ie  functions defined on the Riemannian symmetric space $G/K$, which are invariant under the arithmetic subgroup $G(\ints)$. Following the normalisation of Langlands~\cite{Langlands}, one can define an Eisenstein series for almost all (complex) weights $\lambda$ of $G$ by the formula (see also~\cite{Green:2010kv,Fleig:2012xa,Pioline:2015yea} where also the other statements can be found)
\begin{align}
\label{Eisenstein}
E(\lambda, \cosel) = \sum_{\gamma\in B(\ints)\backslash G(\ints)} e^{\langle \lambda+\rho|H(\gamma \cosel) \rangle}
\end{align}
where $\cosel\in G$ is parametrised by the moduli of the theory in $G/K$.\footnote{The Eisenstein series is by construction \textit{spherical} meaning that $E(\lambda,\cosel k)=E(\lambda,\cosel)$ for all $k\in K$ and can therefore be viewed as a function of the moduli in $G/K$.} The discrete subgroup $G(\ints)$ acts on $G$ by left multiplication $\cosel\mapsto\gamma \cosel$ and the function $H(\cosel)$ picks out the logarithm of the Cartan torus part $a$ in the Iwasawa decomposition $\cosel=nak$ of an element $\cosel\in G$. Therefore $H(\cosel)$ belongs to the Cartan subalgebra $\mathfrak{h}$ of $G$ and can be paired canonically with the weight $\lambda\in \mathfrak{h}^*$.  The element $\rho\in\mathfrak{h}^*$ in~\eqref{Eisenstein} denotes the Weyl vector (half the sum of the positive roots). The series is absolutely convergent for 
\be \langle \mbox{Re}(\lambda)  |  \alpha\rangle > \langle \rho  |  \alpha\rangle \quad \textrm{for all }\alpha  >0 \ , \label{ConvergeL} \ee
for all positive roots, and extends to a meromorphic function of the weight $\lambda$ over $\mathfrak{h}^*$ \cite{Langlands}.  The function $E(\lambda,\cosel)$ is $G(\ints)$ invariant and satisfies the Laplace eigenvalue equation
\begin{align}
\Delta E(\lambda, \cosel) = \frac12\scal{ \langle\lambda | \lambda \rangle - \langle\rho|\rho\rangle}E(\lambda, \cosel).
\end{align}
in terms of the standard bilinear form $\langle\cdot|\cdot\rangle$ on $\mathfrak{h}^*$ which is normalised such that long roots $\alpha$ have length $\langle\alpha|\alpha\rangle =2$.

For the particular series arising in this work it will be convenient to parametrise $\lambda$ as
\begin{align}
\lambda = 2\omega - \rho.
\end{align}
The advantage of this notation is that for {\em maximal parabolic Eisenstein series} the weight $\omega\in\mathfrak{h}^*$ thus defined is proportional to a fundamental weight of $G$. More precisely, we expand $\omega$ on the basis of fundamental weights $\Lambda_i$ ($i=1,\ldots, \textrm{rank}(G)$) as
\begin{align}
\label{Dynklabels}
\omega = \sum_{i=1}^{\dim \mathfrak{h}^*} s_i \Lambda_i.
\end{align}
A maximal parabolic Eisenstein series then has only one non-zero $s_i$. Using~\eqref{Dynklabels} we denote the Eisenstein $E(\lambda,\cosel)$ alternatively by a labelled Dynkin diagram using the numbering conventions of Bourbaki (identical to those of the LiE program~\cite{LiE}). For example, for $SO(5,5)$ of Cartan type $D_5$ we will write
\begin{align}
\label{EisenDynk}
E_{\DSOV{s_1}{s_2}{s_3}{s_4}{s_5}}
\end{align}
and will always suppress the dependence on the coset representative $\cosel\in G$. An example of a maximal parabolic Eisenstein series for $E_{8(8)}$ then would be
\begin{align}
E_{\DEVIII{s}{0}{0}{0}{0}{0}{0}{0}}.
\end{align}
This is defined for almost all complex $s$ (by analytic continuation). The value $s=\frac32$ corresponds to the $R^4$ function~\eqref{R4eqn} and $s=\frac52$ to the $D^4R^4$ correction~\eqref{D4R4eqn}~\cite{Obers:1999um,Green:2010kv}.

We will also sometimes refer to a maximal parabolic Eisenstein series by the representation the relevant fundamental weight $\Lambda_i$ corresponds to. In this terminology
\begin{align}
E_{\DSOV{s}{0}{0}{0}{0}}
\end{align}
will be called a vector Eisenstein series of $SO(5,5)$ (for any $s$) and
\begin{align}
\label{E8adj}
E_{\DEVIII{0}{0}{0}{0}{0}{0}{0}{s}}
\end{align}
an adjoint Eisenstein series of $E_{8(8)}$ (for any $s$). 

We must warn the reader that in this paper we will always consider the Eisenstein series in the Langlands normalisation \eqref{Eisenstein}, whereas one often finds the lattice sum normalisation  that differs by a factor of $2\zeta(2s)$ in the literature.

\subsection{Functional relations and constant term formulas}

Eisenstein series satisfy almost everywhere the functional relation
\begin{align}
\label{Lfunctional} 
E(\lambda,\cosel) = M(w,\lambda) E(w\lambda,\cosel), 
\end{align}
where $w$ is an element of the Weyl group $\mathcal{W}=\mathcal{W}(G)$ and the intertwining coefficient (sometimes also called reflection coefficient) is given by
\begin{align}
\label{intertwiner}
M(w,\lambda) = \prod_{{\substack{\alpha>0 \\ w\alpha<0 }}} \frac{\xi(\langle \lambda |\ \alpha\rangle)}{\xi(\langle \lambda |\ \alpha\rangle+1)},
\end{align}
where the product is over all positive roots $\alpha$ that are mapped to negative roots by the Weyl word $w$. The completed Riemann zeta function 
\begin{align}
\label{complZ}
\xi(s)=\pi^{-s/2} \Gamma(s/2) \zeta(s)
\end{align}
has simple poles at $s=0$ and $s=1$ with residues $-1$ and $+1$, respectively, and vanishes nowhere on the real line. It satisfies the functional relation $\xi(s)=\xi(1-s)$.

Of use to us will also be {\em constant term formulas} that express the integration over some of the variables in the unipotent part $U$ of a parabolic subgroup $P=LU\subset G$ in terms of automorphic functions on the Levi part $L$. In physical terms, the constant term formula expresses the result of averaging out certain axionic moduli and thus projecting to the zero-instanton charge sector for instantons charged under these axions. A parabolic subgroup is the product of an abelian subgroup $GL(1)^{\times n}$ and the semi-simple component of the Levi subgroup. The constant term formula projects onto the perturbative part in the moduli parametrizing this abelian subgroup $GL(1)^{\times n}$, which are generically combinations of the string coupling constant and the radii of the compactification torus. We will label maximal parabolic subgroups (corresponding to $n=1$) by $P_i=L_i U_i$, where $i$ denotes the node of the Dynkin diagram of $G$ that has to be removed to obtain the Dynkin diagram of the Levi subgroup $L_i$ (more precisely that of the semi-simple part $G_i:=[L_i,L_i]$ of $L_i=GL(1)\times [L_i,L_i]$ where $[L_i, L_i]$ denotes the commutator subgroup). 
For a maximal parabolic subgroup $P_i$ the constant formula can be written as~\cite{MW,Green:2010kv,Fleig:2012xa}
\begin{align}
\label{CTmax}
\int_{U_i^1} E(\lambda,u\cosel)du = \sum_{w\in \mathcal{W}_i\backslash \mathcal{W}} M(w,\lambda) e^{\langle (w\lambda+\rho)_{\parallel_i} | H(\cV) \rangle} E^{G_i}((w\lambda)_{\perp_i},\cosel_i).
\end{align}
Let us explain the notation in this formula. The integration domain is $U_i^1:=U_i(\ints)\backslash U_i=(G(\ints)\cap U_i)\backslash U_i$ which is the fundamental domain of the discrete shifts in the unipotent group $U_i(\reals)$ to restrict the integration to a single period. In the simplest case of a one-dimensional unipotent and in a convenient normalisation one has $U_i^1 = \ints\backslash \reals = [0,1)$, the unit interval giving rise to the notational superscript ${}^1$. Since we are averaging over the $U_i$ dependence, the result of the integral can only depend on the variables parametrising $L_i= GL(1)\times G_i$. The dependence on the two factors is separated on the right-hand side, where the dependence on $\cosel_i \in G_i$ is via an Eisenstein series on the group $G_i$ and the dependence on the $GL(1)$ Cartan torus factor is written in terms of the exponential prefactor. A given weight $\lambda$ of $G$ can be decomposed into a component parallel to the fundamental weight $\Lambda_i$ (by orthogonal projection) and remaining components orthogonal to it:
\begin{align}
\lambda = \lambda_{\parallel_i} + \lambda_{\perp_i},\quad\quad \lambda_{\parallel_i} =\frac{\langle \Lambda_i|\lambda\rangle}{\langle \Lambda_i|\Lambda_i\rangle}  \Lambda_i \  .
\end{align}
When $\lambda_{\parallel_i}$ is contracted with an element of the Cartan torus it picks out only the component along the $GL(1)$ factor in the Levi subgroup $L_i$ and therefore the exponential prefactor stands for some power of a variable on $GL(1)$ whose normalisation we will choose to give it an easy physical interpretation. The component $\lambda_{\perp_i}$ is then a combination only of the simple roots of the subgroup $G_i\subset G$ and can therefore be used to define an Eisenstein series on the group $G_i$. The sum in~\eqref{CTmax} is over the quotient of the Weyl groups of $G$ and $G_i$ and the (numerical) coefficient $M(w,\lambda)$ is precisely the intertwiner defined in~\eqref{intertwiner} and hence given by a quotient of completed Riemann zeta functions~\eqref{complZ}. In keeping with our notation we will typically suppress the moduli dependence in the Eisenstein series and label the weight $\lambda$ in terms of its Dynkin diagram representation as in~\eqref{EisenDynk}. (There exists a different constant term formula when $U$ is the maximal unipotent $N$~\cite{Langlands} but we will not need it here.)

As an example for~\eqref{CTmax}, we can consider the following constant term integral for the vector Eisenstein series of $SO(5,5)$\begin{align}
\label{SO55vec}
\int_{U_1^1} E_{\DSOV{s}{0}{0}{0}{0}} = \gs^{-2s} + \gs^{2s-8} \frac{\xi(2s-4)\xi(2s-7)}{\xi(2s)\xi(2s-3)}
+ \gs^{-1} \frac{\xi(2s-1)}{\xi(2s)} E_{\DSOIVS{s-\tfrac12}{0}{0}{0}}\ ,
\end{align}
where $\gs$ denotes a coordinate on the $GL(1)$-part of the Levi subgroup $L_1=GL(1)\times SO(4,4)$ of the maximal parabolic $P_1$ associated with the first (left-most) node of the $SO(5,5)$ diagram. $SO(5,5)$ is the Cremmer--Julia group in $D=6$ space-time dimensions and the constant term formula above corresponds to a string perturbative expansion since it preserves the (chirality preserving) T-duality group $SO(4,4,\ints)$ in $D=6$ and that is why we have labelled the expansion parameter by the six-dimensional string coupling $\gs=\gs_6$. The labelling of the $SO(4,4)$ Dynkin diagram is chosen such that the first node corresponds to the vector representation of the T-duality group, noting that the RR moduli define a Weyl spinor of negative chirality of $Spin(4,4)$.  As is well-known~\cite{Kiritsis:1997em,Obers:1999um,Green:2010kv}, the above Eisenstein series is related to the six-dimensional $R^4$ correction by
\begin{align}
\mathcal{E}_{\gra00}^{\ord{6}}= 2\zeta(3)  E_{\DSOV{\tfrac32}{0}{0}{0}{0}}\  .
\end{align}
For the particular value $s=\tfrac32$, the middle term in the constant term formula~\eqref{SO55vec} vanishes due to the properties of the completed Riemann zeta function and one obtains therefore
\begin{align}
\int_{ U_1^1} \mathcal{E}_{\gra00}^{\ord{6}}= 2\zeta(3) \gs^{-3} +4 \zeta(2)\gs^{-1} E_{\DSOIV{0}{0}{1}{0}} \label{EquaE32}\  ,
\end{align}
corresponding to the correct tree-level and one-loop contributions to the $\tfrac12$-BPS coupling $R^4$.\footnote{As usual, the symmetry is made manifest in Einstein frame, explaining why the powers on the string coupling are shifted from the string frame values $\gs^{-2}$ and $\gs^0$ for tree level and one-loop.}  Manipulations of this kind will be central for evaluating the supergravity divergences in various dimensions that arise from poles in Eisenstein series.

\subsection{Differential equations}
\label{sec:DE}

Eisenstein series are eigenfunctions of all Casimir differential operators on $G/K$ for almost all values of the weight $\lambda\in\mathfrak{h}^*$. To define the differential operators on $G/K$ it is convenient to take a specific representation of the coset representative $\cosel\in G$ in terms of coordinates $\Phi$, as for example the one, $\Phi=(\phi,\sigma)$, associated to the Iwasawa decomposition, such that $\cosel(\Phi)=\cosel(\phi,\sigma)=n(\sigma)a(\phi)k$. One can define in this way the symmetric space connection $Q$ and its  vielbeins $P$ from the components of the Maurer--Cartan form restricted to the Lie algebra $\mathfrak{k}$ of $K$ and its orthogonal complement in $\mathfrak{g}$, \ie 
\be 
\cosel(\Phi)^{-1} d\cosel(\Phi) = Q_\mu(\Phi) d\Phi^\mu+ P_\mu(\Phi) d\Phi^\mu \ , \qquad Q \in \mathfrak{k} \ , \qquad P\in \mathfrak{g}\ominus \mathfrak{k} \ . 
\ee 
The group $K$ defines the structure group of the symmetric space, and one can modify the reference frame by arbitrary functions $k(\Phi)$ such that 
\be 
Q \rightarrow k(\Phi)^{-1} d k(\Phi) +  k(\Phi)^{-1} Q k(\Phi) \ , \qquad P\rightarrow   k(\Phi)^{-1}P  k(\Phi) \ . 
\ee 
The Riemannian metric on the symmetric space   $G/K$ is defined for some appropriately normalised $G$-invariant bilinear form  
\be 
G_{\mu\nu}(\Phi)  = \langle P_\mu(\Phi) ,P_\nu(\Phi)  \rangle \ , 
\ee
and permits to define the inverse vielbeins through its inverse
\be V^\mu \equiv G^{\mu\nu} P_\nu \ . \ee
One defines  the covariant derivative in tangent frame $\cD$,  as the differential operator acting on any tensor function $f_{R_K}(\Phi)$  on  $G/K$ in an arbitrary representation $R_K$ of $K$ and transforming as $f_{R_K} \to \pi_{R_K}(k^{-1} ) f_{R_K}$. The differential operator takes values in the $K$-representation $(\mathfrak{g}\ominus\mathfrak{k})\otimes R_K$ and is defined by
\be 
\cD f_{R_K}(\Phi)  \equiv V^\mu \otimes \scal{ \partial_\mu + \pi_{R_K}(Q_\mu )} f_{R_K}(\Phi)  \ ,  
 \ee
where $\pi_{R_K}(X)$ is the Lie algebra element $X\in\mathfrak{k}$ in the representation $R_K$ acting on  $ f_{R_K}(\Phi)$. 
In particular, for a function defined on $G/K$ such as $\mathcal{E}_{\gra{p}{q}}^{\ord{D}}$, one defines recursively the $n^{\rm th}$ order differential operators
\bea 
\cD f(\Phi) &=& V^\mu\partial_\mu f(\Phi) \CR
\cD\otimes \cD f(\Phi) &=& V^\mu  \otimes \Scal{  \partial_\mu \Scal{ V^\nu \partial_\nu  f(\Phi) } + [ Q_\mu, V^\nu]  \partial_\nu  f(\Phi)} \CR
\cD \otimes \cD\otimes \cD f(\Phi) &=&  V^\mu \otimes  \partial_\mu \Scal{ \cD\otimes \cD f(\Phi) } + V^\mu  \otimes [ Q_\mu , V^\nu]  \otimes \scal{  \partial_\nu \cD f(\Phi)  + [ Q_\nu, \cD f(\Phi)] }\CR
&& \qquad +  V^\mu  \otimes  V^\nu  \otimes \bigl[ Q_\mu ,\scal{  \partial_\nu \cD f(\Phi)  + [ Q_\nu, \cD f(\Phi)] } \bigr] 
\eea
which is valued in the $n^{\rm th}$ tensor power of the Lie algebra component $\mathfrak{g}\ominus\mathfrak{k}$. 

The differential operators can be written in an arbitrary representation $R$ of $\mathfrak{g}$ by writing the coset element $V^\mu$ in the representation $R$. Then powers of the differential operator are mapped to powers in the representation $R$, such that one projects these differential operators valued in the tensor algebra to the enveloping algebra of $\mathfrak{g}$ associated to this representation, and one writes then 
\bea {\bf D}_R f(\Phi) &=& \pi_R(V^\mu) \partial_\mu f(\Phi) \CR
 {\bf D}_R^{\; 2}  f(\Phi) &=& \pi_R(V^\mu) \Scal{  \partial_\mu \Scal{ {\bf D}_R f(\Phi) } + [ \pi_R(Q_\mu), {\bf D}_R f(\Phi)  ]  } \CR
 {\bf D}_R^{\; n+1}f(\Phi)  &=&   \pi_R(V^\mu) \Scal{  \partial_\mu \Scal{ {\bf D}_R^{\; n} f(\Phi) } + [ \pi_R(Q_\mu), {\bf D}_R^{\; n} f(\Phi)  ]  }\ ,  \label{CharacD} \eea
which defines a matrix of differential operators in an explicit matrix representation $R$.  Doing so one considers by construction the restriction of these differential operators to specific irreducible representations of $K$, with the technical advantage that it becomes relatively simple to compute the explicit form of these differential operators in specific parabolic decompositions. 

A generic character   $e^{2 \langle \omega|H(\cosel) \rangle}$ satisfies by construction tensorial differential equations in some irreducible representations of $K$, depending polynomially on the weight vector $\omega$. For $\omega$ such that the bound \eqref{ConvergeL} is satisfied, the associated Eisenstein series is absolutely convergent, and it follows that it also satisfies the same tensorial differential equations for almost all $\omega$ by analytic continuation. For some sub-classes of weight vectors $\omega$, in particular when the latter is proportional to a fundamental weight, \ie  $\omega = s \Lambda_i$, the generating character satisfies stronger differential equations that can often be rewritten as characteristic equations in a given representation $R$, \ie
\be \cP_{\omega,R}({\bf D}_R) e^{2 \langle \omega|H(\cosel) \rangle} = 0 \ , \ee
for a polynomial  $\cP_{\omega,R}({\bf D}_R)$ in the covariant derivative ${\bf D}_R$ that depends polynomially on the weight vector $\omega$ (see \cite{Bossard:2015uga} for some examples). Whenever this sub-class defines a domain intersecting with the domain of convergence of the corresponding Eisenstein series $ E(\lambda,\cosel)$ at $\lambda = 2 \omega - \rho$ \eqref{ConvergeL}, it follows that the latter satisfies 
\be 
\cP_{\omega,R}({\bf D}_R) E(\lambda,\cosel) = 0 \ . 
\ee
Because $E(\lambda,\cosel) $ is analytic in $\omega$ for almost all $\omega$ and the differential operator  $\cP_{\omega,R}({\bf D}_R)$ is analytic in $\omega$, this equation is then satisfied in general for  $\omega$. 

It may also happen that a generating character $e^{2 \langle \omega|H(\cosel) \rangle}$ satisfies a stronger characteristic equation for an isolated $\omega$, for which the corresponding Eisenstein series is not absolutely convergent. Then one cannot directly conclude that the Eisenstein series satisfies itself this stronger characteristic equation. In particular, when the  differential constraint on the character is associated to a nilpotent orbit that is special in the sense of \cite{Lusztig} the corresponding Eisenstein series will also satisfy a weaker constraint associated with a larger special orbit~\cite{BarbaschVogan,Jiang:2014}.

 Let us for this purpose consider the example of $E_{7(7)}$ Eisenstein series. It was computed in \cite{Bossard:2015uga} that
 \bea {\bf D}_{56}^{\; 3}  e^{2s \langle \Lambda_1 |H( \cosel) \rangle} &=&\Scal{ \frac{s(2s-17)}{2} +6  } {\bf D}_{56} e^{2s \langle \Lambda_1 |H(\cosel) \rangle} \ , \CR
 {\bf D}_{133}^{\; 3}  e^{2s \langle \Lambda_7 |H(\cosel) \rangle} &=&s(s-9)   {\bf D}_{133} e^{2s \langle \Lambda_7 |H(\cosel) \rangle}  \ , \eea
and because the corresponding Eisenstein series are  respectively absolutely  convergent for $s>\frac{17}{2}$ and $s>9$, one concludes that for almost all $s$ (\ie away from the poles)
 \begin{subequations}
 \begin{align}
 {\bf D}_{56}^{\; 3} E_{\DEVII{s}{0}{0}{0}{0}{0}{0}} &= \Scal{ \frac{s(2s-17)}{2} +6  }  {\bf D}_{56}  E_{\DEVII{s}{0}{0}{0}{0}{0}{0}}  \ , \\
 \label{TDfun}
 {\bf D}_{133}^{\; 3}  E_{\DEVII{0}{0}{0}{0}{0}{0}{s}}  &= s(s-9)   {\bf D}_{133}  E_{\DEVII{0}{0}{0}{0}{0}{0}{s}}  \ . 
 \end{align}
 \end{subequations}
In type II string theory, the $\nabla^4 R^4$ threshold function is conjectured to be $\zeta(5) E{\DEVII{\sfrac{5}{2}}{0}{0}{0}{0}{0}{0}}$ \cite{Green:2010kv}, and must satisfy by supersymmetry \cite{Bossard:2014lra,Bossard:2014aea}  the two equations 
\be 
{\bf D}_{56}^{\; 3} E_{\DEVII{\sfrac{5}{2}}{0}{0}{0}{0}{0}{0}} =-9 \,   {\bf D}_{56}  E_{\DEVII{\sfrac{5}{2}}{0}{0}{0}{0}{0}{0}}  \ , \qquad  {\bf D}_{133}^{\; 3}  E_{\DEVII{\sfrac{5}{2}}{0}{0}{0}{0}{0}{0}}  =-20 \,   {\bf D}_{133}  E_{\DEVII{\sfrac{5}{2}}{0}{0}{0}{0}{0}{0}}  \ . 
\ee
The first equation is obviously satisfied by $\zeta(5) E{\DEVII{\sfrac{5}{2}}{0}{0}{0}{0}{0}{0}}$. For the second equation, one has to use that,  because of the Langlands functional identity \eqref{Lfunctional}
\be 
\zeta(5) E_{\DEVII{\sfrac{5}{2}}{0}{0}{0}{0}{0}{0}} = \frac{8\zeta(8)}{15\pi} E_{\DEVII{0}{0}{0}{0}{0}{0}{4}}  \ , 
\ee
this function is a special case of both the adjoint and the fundamental Eisenstein series. From equation~\eqref{TDfun} for the fundamental Eisenstein series one then sees that both supersymmetry constraints are fulfilled. Moreover, one can in this way understand that this function admits a wave-front set associated to the next-to-minimal nilpotent orbit of $E_7$ \cite{Green:2011vz}. 

By supersymmetry, the type II string theory $R^4$ threshold function must satisfy the stronger differential equation \cite{Bossard:2014lra}  
\begin{align}
{\bf D}_{56}^{\; 2} \mathcal{E}_{\gra00}^{\ord{4}}  =- \frac{9}{2} \mathds{1}_{56}   \mathcal{E}_{\gra00}^{\ord{4}}
\end{align}
and the conjectured solution $\mathcal{E}_{\gra00}^{\ord{4}}=2\zeta(3) E{\DEVII{\sfrac{3}{2}}{0}{0}{0}{0}{0}{0}}$ \cite{Obers:1999um,Green:2010kv} indeed solves this constraint~\cite{Bossard:2014lra}.

However, one must in general be careful when the Eisenstein series is outside the domain of absolute convergence. We will see in section~\ref{sec:FourierD5} that even though the character $e^{3\langle \Lambda_2|H(\cosel)\rangle}$ of the adjoint series of $SO(5,5)$ satisfies a certain stronger constraint the associated Eisenstein series (and its Fourier coefficients) do not. This also happens for the adjoint series of $E_{7(7)}$ and the character $e^{8 \langle \Lambda_1 |H(\cosel) \rangle}$ that satisfies an additional  quartic differential equation in the $[2,0,0,0,0,0,2]$ of $SU(8)$, while the corresponding adjoint function at $s=4$ violates this constraint.  In appendix~\ref{E8Diff} we show the same for the adjoint $E_{8(8)}$ series.

\section{Supersymmetry constraints on $\cE_\gra01$}
\label{sec:SUSYE01}

It was shown in \cite{Bossard:2015uga} that the $\nabla^6 R^4$ threshold function $\mathcal{E}_\gra{0}{1}^{\ord{D}}$ decomposes into the sum of two distinct functions associated to two different supersymmetry invariants for $D>3$. The two invariants are distinguished by higher point R-symmetry violating couplings. One function satisfies a homogeneous differential equation and is conjectured to be an Eisenstein series, whereas the other satisfies an inhomogeneous equation as was first argued in \cite{Green:2005ba}. For instance, in six dimensions, $E_5\cong SO(5,5)$ and $K\cong SO(5)\times SO(5)$, and the threshold function $\mathcal{E}_\gra{0}{1}$ decomposes as
\be 
\label{E01D6}
\mathcal{E}_\gra{0}{1}  = \mathcal{F}_\gra{0}{1}  + \frac{4\zeta(6)\xi(8)}{27\xi(4)} \hat{E}_{\mbox{\DSOV{0}0004}}\ . 
\ee
(The hat here denotes a regularised spinor Eisenstein series that will be defined in~\eqref{D5s4reg} below.)
The supersymmetry analysis only constrains the function appearing in the Wilsonian action, and one must consider possible anomalous corrections to the corresponding differential equation whenever there are logarithmic divergences in the theory. It turns out in particular that the four-graviton amplitude diverges at 3-loop in six dimensions \cite{Bern:2008pv}, and  supersymmetry therefore only constrains the function  $\mathcal{F}_\gra{0}{1}$ to satisfy 
\be 
\Delta \mathcal{F}_\gra{0}{1} = - \Scal{ 2\zeta(3) E_{\DSOV{\stfrac{3}{2}}{0}{0}{0}{0}}}^2  + \frac{70}{3} c_1 \zeta(3) \ , \label{AnomaLaplace} 
\ee
where $c_1$ is a constant yet to be determined, and \cite{Bossard:2015uga}
\begin{subequations}
\label{EquaF1}
\begin{align}
 {\bf D}_{16}^{3} \mathcal{F}_\gra{0}{1} &=  \frac{3}{4} {\bf D}_{16} \mathcal{F}_\gra{0}{1}  -2 \zeta(3)^2  E_{\DSOV{\stfrac{3}{2}}{0}{0}{0}{0}}  {\bf D}_{16} E_{\DSOV{\stfrac{3}{2}}{0}{0}{0}{0}}   \ ,  \\
 {\bf D}_{10}^{3} \mathcal{F}_\gra{0}{1} &= \frac{3}{2} {\bf D}_{10} \mathcal{F}_\gra{0}{1} -2 \zeta(3)^2  E_{\DSOV{\stfrac{3}{2}}{0}{0}{0}{0}}  {\bf D}_{10} E_{\DSOV{\stfrac{3}{2}}{0}{0}{0}{0}}  \ .
 \end{align}
 \end{subequations}
Here, ${\bf D}_{16}$ refers to the covariant derivative valued in the (chiral) spinor representation, and  ${\bf D}_{10}$ to the covariant derivative valued in the vector representation, according to the notation introduced in the preceding section. One can for example write ${\bf D}_{16}$ and ${\bf D}_{10}$ as  explicit matrices of differential operators as follows 
\be 
{\bf D}_{10} = \begin{pmatrix} 0 & \cD_{a\hat{b}} \\ \cD_{b\hat{a}} & 0 \end{pmatrix}  \ , \qquad {\bf D}_{16} =  \frac{1}{2} \cD_{a\hat{b}} \gamma^a \gamma^{\hat{b}}  \    , 
\ee 
where $\cD_{a\hat{b}}$ is the covariant derivative as a  $({\bf 5},{\bf 5})$ tensor of $SO(5)\times SO(5)$, and we label the vector indices of the first factor by $a$ (ranging from $1$ to $5$) and those of the second factor by $\hat{a}$. $\gamma^a,\, \gamma^{\hat{a}}$ are the $Spin(5,5)$ gamma matrices in a (fixed) Majorana--Weyl representation and $SO(5)$ indices are raised and lowered with the flat metric.

It is important to note that equations \eqref{EquaF1} are invariant with respect to the exchange of chirality, and read in $SO(5)\times SO(5)$ covariant notations 
\begin{subequations}
\begin{align} 
\varepsilon^{abcde} \varepsilon_{\hat{a}\hat{b}\hat{c}\hat{d}\hat{e}} \cD_{a}{}^{\hat{a}} \cD_{b}{}^{\hat{b}} \cD_{c}{}^{\hat{c}}  \mathcal{F}_\gra{0}{1} &= 0 \ ,  \\
\cD_{a\hat{c}} \cD^{d\hat{c}} \cD_{d\hat{b}}\mathcal{F}_\gra{0}{1}  &= \frac{3}{2} \cD_{a\hat{b}} \mathcal{F}_\gra{0}{1}  -2 \zeta(3)^2  E_{\DSOV{\stfrac{3}{2}}{0}{0}{0}{0}} \cD_{a\hat{b}}  E_{\DSOV{\stfrac{3}{2}}{0}{0}{0}{0}}  \ .\label{InhomoD5}  
\end{align}
\end{subequations}
On the other hand,  one computes using\footnote{The regularised function is only defined up to an arbitrary  additional constant, which is associated to the ambiguity in defining the separation between the local and the non-local components of the effective action in the presence of logarithm terms.}
\be   
\label{D5s4reg}
\hat{E}_{\mbox{\DSOV{0}0004}} = \lim_{\epsilon\rightarrow 0} \Scal{ {E}_{\mbox{\DSOV{0}000{4\mbox{+}\epsilon}}} - \frac{\xi(3)}{\xi(6)\xi(8)} \frac{1}{2\epsilon} } 
\ee
that \cite{Bossard:2015uga}
\begin{subequations}
\begin{align}
\Delta  \hat{E}_{\mbox{\DSOV{0}0004}} &= 5 \frac{\xi(3)}{\xi(6)\xi(8)} \ , \\
 \cD_{[a}{}^{[\hat{a}} \cD_{b}{}^{\hat{b}} \cD_{c]}{}^{\hat{c}]} \,  \hat{E}_{\mbox{\DSOV{0}0004}} &=  - \frac{1}{12} \varepsilon_{abcde} \varepsilon^{\hat{a}\hat{b}\hat{c}\hat{d}\hat{e}}  \cD^d{}_{\hat{d}} \cD^e{}_{\hat{e}} \hat{E}_{\mbox{\DSOV{0}0004}} \ , \label{HomoD5}
\end{align}
\end{subequations}
which is not invariant under the parity transformation in the T-duality group. 
 
 \subsection{Solution using string perturbation theory}
 
 We will now construct the solution to the tensorial differential equation~\eqref{EquaF1} using string perturbation theory. At the end of the derivation we will argue that the differential equation \eqref{EquaF1} does not admit cusp form solutions, and the method provides the full non-perturbative solution.

According to string perturbation theory, the function $\mathcal{F}_\gra{0}{1} $ decomposes as
\be 
\mathcal{F}_\gra{0}{1} = e^{-6\phi} \sum_{\ell =0}^\infty e^{2\ell \phi} \mathcal{F}^\ord{\ell}_\gra{0}{1} + \frac{10\zeta(3)}{3} c_2 \, \phi +  \mathcal{O}(e^{-e^{-\phi}}) \ , 
\ee
where $e^{\phi} = \gs$ is the string theory effective coupling constant in six dimensions, and the additional term linear in $\phi$ must be added to take into account the non-analyticity of the threshold function due to the 3-loop divergence \cite{Green:2010sp}. To solve these equations in the  string perturbation  theory limit, we need the explicit decomposition of these differential operators acting on a function $\cF$ of the dilaton $\phi$ and the $SO(4,4)$ scalar fields\footnote{Here $\frac{1}{4}  \partial_\phi + {\bf D}_{8}$ is understood to be $\frac{1}{4}  \partial_\phi \mathds{1}_{8} + {\bf D}_{8}$. We will never write  explicitly the identity matrices in the following.}
\be 
{\bf D}_{10} \cF = \left( \begin{array}{ccc} \ \frac{1}{2} \partial_\phi \ & 0 & 0 \\
0 & \ {\bf D}_{8_{\rm \rm a}}  \ & 0 \\
0 & 0 &  - \frac{1}{2} \partial_\phi \ \end{array} \right)  \cF \ , \qquad {\bf D}_{16} \cF = \left( \begin{array}{cc} \ \frac{1}{4}  \partial_\phi + {\bf D}_{8}  \ & 0  \\
 0 &  - \frac{1}{4} \partial_\phi  + {\bf D}_{8_{\rm  c}} \ \end{array} \right)  \cF\ . 
\ee
Note that we use the embedding of $SO(4,4)$ as the T-duality group in string theory, referring to the property that the $16$ vector fields in six dimensions are associated to $8$ NS and $8$ RR fields (respectively in the ${\bf 8}$ and the ${\bf 8}_{\rm  c}$ of $Spin(4,4)$). One computes that the second order differential operator defined as in \eqref{CharacD} is
\bea 
{\bf D}_{10}^{\; 2} \cF &=& \left( \begin{array}{ccc} \ \frac{1}{4} \partial^{\; 2}_\phi + \partial_\phi  \ & 0 & 0 \\
0 & \ {\bf D}_{8_{\rm \rm a}}^{\; 2} + \frac{1}{4} \partial_\phi   \ & 0 \\
0 & 0 &  \frac{1}{4} \partial^{\; 2}_\phi + \partial_\phi  \ \end{array} \right)  \cF \ , \CR
 {\bf D}_{16}^{\; 2} \cF &=& \left( \begin{array}{cc} \ \frac{1}{16}  \partial^{\; 2}_\phi  + \frac{1}{2} \partial_\phi + \frac{1}{2} \scal{ \partial_\phi + 1}  {\bf D}_{8} + {\bf D}^{\; 2}_{8}  \hspace{-10mm}  & 0  \\
 0 & \hspace{-10mm}  \frac{1}{16}  \partial^{\; 2}_\phi  + \frac{1}{2} \partial_\phi - \frac{1}{2} \scal{ \partial_\phi + 1}  {\bf D}_{8_{\rm  c}} + {\bf D}^{\; 2}_{8_{\rm  c}}   \ \end{array} \right)  \cF\ ,\eea
and finally
\bea 
{\bf D}_{10}^{\; 3} \cF &=& \left( \begin{array}{ccc} \ \frac{1}{8} \partial^{\; 3}_\phi +\partial_\phi^{\; 2}  + \frac{3}{2} \partial_\phi- \frac{1}{4} \Delta    & 0 & 0 \\
0 &  {\bf D}_{8_{\rm \rm a}}^{\; 3} + \frac{1}{4} \partial_\phi {\bf D}_{8_{\rm \rm a}}    & 0 \\
0 & 0 &  -\frac{1}{8} \partial^{\; 3}_\phi -\partial_\phi^{\; 2}  - \frac{3}{2} \partial_\phi+\frac{1}{4} \Delta   \ \end{array} \right)  \cF \ , \\
 {\bf D}_{16}^{\; 3} \cF &=&\hspace{-2mm}  \left( \begin{array}{cc}  \hspace{-1mm} \frac{1}{64}  \partial^{\; 3}_\phi  + \frac{1}{8} \partial_\phi^{\; 2} - \frac{1}{8} \Delta +  \scal{ \frac{3}{16} \partial_\phi^{\; 2} + \frac{11}{8} \partial_\phi + \frac{3}{4} }  {\bf D}_{8} + \frac{3}{4}\scal{ \partial_\phi + 2}  {\bf D}^{\; 2}_{8}   + {\bf D}_{8}^{\; 3} \hspace{-10mm}  & 0  \\
 \hspace{-80mm} 0 & \hspace{-90mm} - \frac{1}{64}  \partial^{\; 3}_\phi  - \frac{1}{8} \partial_\phi^{\; 2} + \frac{1}{8} \Delta +  \scal{ \frac{3}{16} \partial_\phi^{\; 2} + \frac{11}{8} \partial_\phi + \frac{3}{4} }  {\bf D}_{8_{\rm c}} - \frac{3}{4}\scal{ \partial_\phi + 2}  {\bf D}^{\; 2}_{8_{\rm c}}   + {\bf D}_{8_{\rm c}}^{\; 3}    \hspace{-1mm}\end{array} \right)  \cF\ . \nn 
 \eea
We recall that the square notation is a short-hand notation for the definition \eqref{CharacD}, explaining the additional lower order differential operator contributions associated with the terms involving the connection. By construction 
\be 
\Delta_{D_5} \cF = \tr {\bf D}_{10}^{\; 2} \cF = \frac{1}{2} \tr {\bf D}_{16}^{\; 2} \cF  = \Scal{ \Delta_{D_4} + \frac{1}{2} \partial_\phi^{\; 2} +  4 \partial_\phi } \cF \ . \label{PertuLaplace} 
\ee
To solve these differential equations, we will make use of the particular solutions 
\be
\begin{split}  {\bf D}_8^{\; 3}  E_{\DSOIV{s}{0}{0}{0}} &= \scal{ s(s-3) + \tfrac{3}{2}} {\bf D}_8 E_{\DSOIV{s}{0}{0}{0}} \ , \\
 {\bf D}_{8_{\rm c}}^{\; 2}  E_{\DSOIV{s}{0}{0}{0}} &= \frac{s(s-3)}{4}  E_{\DSOIV{s}{0}{0}{0}}\ ,  \\
  {\bf D}_{8_{\rm a}}^{\; 2}  E_{\DSOIV{s}{0}{0}{0}} &= \frac{s(s-3)}{4}  E_{\DSOIV{s}{0}{0}{0}}\ ,  \\
 \end{split}\quad\begin{split} {\bf D}_{8}^{\; 2}  E_{\DSOIV{0}{0}{0}{s}} &= \frac{s(s-3)}{4}  E_{\DSOIV{0}{0}{0}{s}}\ ,  \\
   {\bf D}_{8_{\rm c}}^{\; 3}  E_{\DSOIV{0}{0}{0}{s}} &= \scal{ s(s-3) + \tfrac{3}{2}} {\bf D}_{8_c} E_{\DSOIV{0}{0}{0}{s}} \ , \\
  {\bf D}_{8_{\rm a}}^{\; 2}  E_{\DSOIV{0}{0}{0}{s}} &= \frac{s(s-3)}{4}  E_{\DSOIV{0}{0}{0}{s}}\ ,  \\
\end{split} \label{D4Eq} 
\ee
and equivalently for $E_{\DSOIV{0}{0}{s}{0}} $. 

We will now analyse the differential equations order by order in string perturbation theory, \ie for the various $\cF_{\gra01}^{\ord{\ell}}$.
Using equation \eqref{EquaE32} with $\gs = e^\phi$, one can now compute 
 \bea \Scal{ {\bf D}_{10}^{\; 3} - \frac{3}{2} {\bf D}_{10}}\frac{2 \zeta(3)^2}{3} e^{-6\phi} &=& 3 \zeta(3)^2\left( \begin{array}{ccc} \ 1\ & 0 & 0 \\
0 & \ 0 \ & 0 \\
0 & 0 &  - 1 \ \end{array} \right)  e^{-6\phi} =  - 2 \zeta(3)^2 e^{-3\phi} {\bf D}_{10} e^{-3\phi} \ , \CR
\Scal{ {\bf D}_{16}^{\; 3} - \frac{3}{4} {\bf D}_{16}}\frac{2 \zeta(3)^2}{3} e^{-6\phi} &=& \frac{3}{2} \zeta(3)^2\left( \begin{array}{cc} \ 1\ & 0  \\ 0 &  - 1 \ \end{array} \right)  e^{-6\phi} =  - 2 \zeta(3)^2 e^{-3\phi} {\bf D}_{16} e^{-3\phi} \ , \eea
such that 
\be 
\mathcal{F}^\ord{0}_\gra{0}{1}  = \frac{2 \zeta(3)^2}{3}  \ . 
\ee
Similarly, one computes that 
\bea 
\Scal{ {\bf D}_{10}^{\; 3} - \frac{3}{2} {\bf D}_{10}}\frac{4 \zeta(2)\zeta(3)}{3} e^{-4\phi} E_{\DSOIVS{0}{0}{0}{1}} &=&4 \zeta(2)\zeta(3) \left( \begin{array}{ccc} \ 2\ & 0 & 0 \\
0 & - {\bf D}_{8_{\rm \rm a}}  \ & 0 \\
0 & 0 &  - 2 \ \end{array} \right) e^{-4\phi} E_{\DSOIVS{0}{0}{0}{1}} \CR
&=&  - 4  \zeta(2)\zeta(3)  {\bf D}_{10} e^{-4\phi} E_{\DSOIVS{0}{0}{0}{1}}  \ , \CR
\Scal{ {\bf D}_{16}^{\; 3} - \frac{3}{4} {\bf D}_{16}} \frac{4 \zeta(2)\zeta(3)}{3} e^{-4\phi} E_{\DSOIVS{0}{0}{0}{1}} &=& 4 \zeta(2)\zeta(3) \left( \begin{array}{cc} \ 1-{\bf D}_{8}  & 0 \   \\ 0 &  - 1-{\bf D}_{8_{\rm  c}}  \ \end{array} \right)  e^{-4\phi} E_{\DSOIVS{0}{0}{0}{1}} \CR
&=&- 4  \zeta(2)\zeta(3)  {\bf D}_{16} e^{-4\phi} E_{\DSOIVS{0}{0}{0}{1}} \ , 
\eea
such that 
\be 
\mathcal{F}^\ord{1}_\gra{0}{1}  = \frac{4 \zeta(2)\zeta(3)}{3}  E_{\DSOIVS{0}{0}{0}{1}}  \ .
 \ee
Note moreover that there is no homogeneous solution to these differential equations with the corresponding factor of the dilaton, such that these solutions are unique at these orders. The 2-loop contribution satisfies the more complicated equations 
 \bea 
 \Scal{ {\bf D}_{10}^{\; 3} - \frac{3}{2} {\bf D}_{10}} e^{-2\phi} \mathcal{F}^\ord{2}_\gra{0}{1} &=& \left( \begin{array}{ccc} \ \frac{3}{2} - \frac{1}{4} \Delta    & 0 & 0 \\
0 &  {\bf D}_{8_{\rm \rm a}}^{\; 3} -2 {\bf D}_{8_{\rm \rm a}}    & 0 \\
0 & 0 &  - \frac{3}{2} +\frac{1}{4} \Delta   \ \end{array} \right)   e^{-2\phi} \mathcal{F}^\ord{2}_\gra{0}{1} \CR
=  -   {\bf D}_{10} \Scal{2  \zeta(2) e^{-\phi} E_{\DSOIVS{0}{0}{0}{1}}}^2 &=& \left( \begin{array}{ccc} \ 1\ & 0 & 0 \\
0 & - {\bf D}_{8_{\rm \rm a}}  \ & 0 \\
0 & 0 &  - 1 \ \end{array} \right) \Scal{2  \zeta(2) e^{-\phi} E_{\DSOIVS{0}{0}{0}{1}}}^2
\eea
and
\bea
\Scal{ {\bf D}_{16}^{\; 3} - \frac{3}{4} {\bf D}_{16}} e^{-2\phi} \mathcal{F}^\ord{2}_\gra{0}{1} &=& \left( \begin{array}{cc} \ \frac{3}{4}- \frac{1}{8} \Delta    + {\bf D}_{8}^{\; 3}-2  {\bf D}_{8} \hspace{-10mm}  & 0  \\
 0 &- \frac{3}{4}+ \frac{1}{8} \Delta    + {\bf D}_{8_{\rm  c}}^{\; 3}-2  {\bf D}_{8_{\rm  c}}    \ \end{array} \right)  e^{-2\phi} \mathcal{F}^\ord{2}_\gra{0}{1}  \CR
=  -   {\bf D}_{16} \Scal{2  \zeta(2) e^{-\phi} E_{\DSOIVS{0}{0}{0}{1}}}^2 &=&  \left( \begin{array}{cc} \ \frac{1}{2} -{\bf D}_{8}  & 0 \   \\ 0 &  - \frac{1}{2}-{\bf D}_{8_{\rm  c}}  \ \end{array} \right)   \Scal{2  \zeta(2) e^{-\phi} E_{\DSOIVS{0}{0}{0}{1}}}^2 \ , 
\eea
 from which one deduces that 
 \be \Delta \mathcal{F}^\ord{2}_\gra{0}{1}  = 6  \mathcal{F}^\ord{2}_\gra{0}{1} -   \Scal{4  \zeta(2) E_{\DSOIVS{0}{0}{0}{1}}}^2\ , \qquad \scal{ {\bf D}_{8_i}^{\; 3} - 2{\bf D}_{8_i} }\mathcal{F}^\ord{2}_\gra{0}{1}  = - {\bf D}_{8_i}   \Scal{2  \zeta(2) E_{\DSOIVS{0}{0}{0}{1}}}^2 \ ,  \ee
 where ${\bf 8}_i$ stands for the three fundamental representations of $Spin(4,4)$, the vector and the two Weyl spinor representations. It is important to note that this tensorial equation is triality invariant since
 \be 
E_{\DSOIV{0}{0}{0}{1}} = E_{\DSOIV{1}{0}{0}{0}} = E_{\DSOIV{0}{0}{1}{0}}\ . 
\ee
This property is crucial in the conjecture proposed in \cite{Pioline:2015yea} that the function $\mathcal{F}_\gra{0}{1}$ is triality related to the genus two $\nabla^6 R^4$ threshold function in five dimensions. The differential equation for $\cF_{\gra01}^{\ord{2}}$ admits an $SO(4,4,\mathds{Z})$ invariant homogeneous solution, such that the differential equation only determines the correct function up to 
 \be  
 \mathcal{F}^\ord{2}_\gra{0}{1} =  \mathcal{F}^\ord{2}_{\gra{0}{1}\, {\rm \scriptscriptstyle part}} + \frac{8 \pi^6 \zeta(5)^2}{496125 \zeta(7)} c_3 E_{\DSOIV{0}{3}{0}{0}} \ . 
 \ee
Let us now derive the 3-loop contribution. One cannot disentangle the $\phi$ independent function $\cF_\gra{0}{1}^\ord{3}$ from the logarithm term linear in $\phi$ in the differential equation, so we consider 
 \bea && \Scal{ {\bf D}_{10}^{\; 3} - \frac{3}{2} {\bf D}_{10}} \Scal{  \mathcal{F}^\ord{3}_\gra{0}{1}+ \frac{10\zeta(3)}{3} c_2 \, \phi } \CR
 &=& \left( \begin{array}{ccc} \  \frac{5\zeta(3)}{2} c_2 - \frac{1}{4} \Delta    & 0 & 0 \\
0 &  {\bf D}_{8_{\rm \rm a}}^{\; 3} -\frac{3}{2} {\bf D}_{8_{\rm \rm a}}    & 0 \\
0 & 0 &   -\frac{5\zeta(3)}{2} c_2 +\frac{1}{4} \Delta   \ \end{array} \right)    \mathcal{F}^\ord{3}_\gra{0}{1}  = 0 \CR
&& \Scal{ {\bf D}_{16}^{\; 3} - \frac{3}{4} {\bf D}_{16}} \Scal{  \mathcal{F}^\ord{3}_\gra{0}{1}+ \frac{10\zeta(3)}{3} c_2 \, \phi }  \CR
 &=& \left( \begin{array}{cc} \  -\frac{5\zeta(3)}{8} c_2- \frac{1}{8} \Delta + \frac{3}{2}   {\bf D}^{\; 2}_{8}   + {\bf D}_{8}^{\; 3}   & 0  \\
 0 & \frac{5\zeta(3)}{8} c_2+ \frac{1}{8} \Delta - \frac{3}{2}   {\bf D}^{\; 2}_{8_{\rm  c}}   + {\bf D}_{8_{\rm  c}}^{\; 3}     \ \end{array} \right) \mathcal{F}^\ord{3}_\gra{0}{1}  = 0 \ , \eea
such that 
\be {\bf D}_{8_{\rm \rm a}}^{\; 3} \mathcal{F}^\ord{3}_\gra{0}{1}  = \frac{3}{2} {\bf D}_{8_{\rm \rm a}} \mathcal{F}^\ord{3}_\gra{0}{1} \ , \quad {\bf D}_{8}^{\; 2}  \mathcal{F}^\ord{3}_\gra{0}{1} = \frac{5\zeta(3)}{4} c_2 \ , \quad {\bf D}_{8_{\rm  c}}^{\; 2}  \mathcal{F}^\ord{3}_\gra{0}{1} = \frac{5\zeta(3)}{4} c_2\ . \ee
 These differential equations are solved by the regularised Eisenstein series 
 \be \hat{E}_{\DSOIVS{3}{0}{0}{0}} = \lim_{\epsilon \rightarrow 0} \Scal{  {E}_{\DSOIVS{3\mbox{+}\epsilon}{0}{0}{0}} - \frac{\xi(3)}{\xi(4)\xi(6)} \frac{1}{2\epsilon}  } \ , \ee
 which by construction \eqref{D4Eq} satisfies  
 \be  {\bf D}_{8_{\rm \rm a}}^{\; 3} \hat{E}_{\DSOIVS{3}{0}{0}{0}}   = \frac{3}{2} {\bf D}_{8_{\rm \rm a}} \hat{E}_{\DSOIVS{3}{0}{0}{0}}  \ , \quad {\bf D}_{8}^{\; 2}  \hat{E}_{\DSOIVS{3}{0}{0}{0}}  =  \frac{3}{8} \frac{\xi(3)}{\xi(4)\xi(6)}   \ , \quad {\bf D}_{8_{\rm  c}}^{\; 2}  \hat{E}_{\DSOIVS{3}{0}{0}{0}}  = \frac{3}{8} \frac{\xi(3)}{\xi(4)\xi(6)}  \ .\ee
Assuming that there is no cusp form satisfying these differential equations, one can argue that this is the unique $SO(4,4,\mathds{Z})$ solution with a sufficiently fast fall-off at the boundary of moduli space. To show this we use the property that the differential equations determine the eigenvalues of all the Casimir operators. It follows that the general solution on the maximal torus (\ie the infinitesimal quasi-character fixed by the weight $\lambda$ in~\eqref{Eisenstein}) is uniquely determined up to Weyl reflections, and so is the general Eisenstein series solution. The spectral decomposition of automorphic forms is into a continuous part (given by Eisenstein series) and a discrete part (corresponding to cusp forms and residues of Eisenstein series)~\cite{Terras}. One can verify that there is no residual spectrum for the weight $\lambda$ defined by the above tensorial differential equations.  Therefore the assumption on the absence of cusp forms implies that the $SO(4,4,\ints)$ Eisenstein series solution is the unique solution. The existence of a cusp form satisfying these differential equations would not affect our conclusions as the cusp form would not contribute to the perturbative series on which we base our analysis.
 
 As we will discuss in the following, comparison with higher dimensional string theory computations allow to determine  $ \mathcal{F}^\ord{3}_\gra{0}{1} $  to be \be \mathcal{F}^\ord{3}_\gra{0}{1}  = \frac{4\zeta(6)}{27}  \hat{E}_{\DSOIVS{3}{0}{0}{0}} \ , \ee
such that $c_2=1$. Using this function, one computes that 
 \be \Delta \Scal{  \frac{4\zeta(6)}{27}  \hat{E}_{\DSOIVS{3}{0}{0}{0}}  +  \frac{10\zeta(3)}{3}  \, \phi } = \frac{70\zeta(3)}{3} \ ,  \label{SolTeDif}\ee
 and therefore $c_1=1$. 
 
By arguments similar to above,  there is no solution to the differential equation at higher order in string perturbation theory. This establishes the expected non-renormalisation theorem that the function $\cE_\gra{0}{1}$ is exact at 3-loop in perturbation theory. Combining these results one deduces that
 \begin{multline}  \cE_\gra{0}{1}  = \frac{2 \zeta(3)^2}{3} e^{-6\phi}  +e^{-4\phi}\Scal{  \frac{4 \zeta(2)\zeta(3)}{3}  E_{\DSOIVS{0}{0}{0}{1}}  + \frac{16 \zeta(8)}{189} E_{\DSOIVS{0}{0}{0}{4}}}   + e^{-2\phi} \cF_\gra{0}{1}^\ord{2}   \\ +  \frac{4\zeta(6)}{27}  \Scal{ \hat{E}_{\DSOIVS{3}{0}{0}{0}} +    \hat{E}_{\DSOIVS{0}{0}{3}{0}}} + 5 \zeta(3) \phi + \mathcal{O}(e^{-e^{-\phi}})  \ , \end{multline}
 in perfect agreement with \cite{Pioline:2015yea}. 
 
We would like now to argue that the differential equation \eqref{EquaF1} does not admit cusp form solutions, and therefore determines  the complete non-perturbative function uniquely. Cusp forms are square integrable functions $\cE$ on the modular domain $\cF_G \cong G(\mathds{Z})\backslash G/K$ and eigenfunctions of the Laplace operator. Due to their cuspidal nature they have strictly negative Laplace eigenvalue since
 \be 
 \int_{\cF_G} d\mu \,  \cE\Delta \cE = - \int_{\cF_G} d\mu\,  |\nabla \cE |^2 <0 \ .
 \ee
Therefore, there can be no cusp form solution to equation \eqref{EquaF1}. 
 
 Let us summarise what we have achieved through this computation. Consistency with higher dimensional results permits to determine that the 3-loop contribution   $ \mathcal{F}^\ord{3}_\gra{0}{1} $ is equal to $  \frac{4\zeta(6)}{27}  \hat{E}{\DSOIVS{3}{0}{0}{0}}  +  c_2\frac{10\zeta(3)}{3}  \, \phi$, and we have shown that the tensorial differential equation \eqref{EquaF1} is strong enough to determine $c_2$ and the anomalous source term in the Laplace equation, \ie $c_1$. 
 
 \subsection{Alternative derivation of non-analytic terms}
 
We will now argue that one can alternatively use the regularised Eisenstein series $\hat{E}{\DSOV{0}{\stfrac{7}{2}}{0}{0}{0}} $ to probe these properties, although the latter does not appear in the complete threshold function $\mathcal{F}^\ord{3}_\gra{0}{1}$.  The adjoint Eisenstein series satisfies in general that 
 \bea 
 \label{eq:DEadjD5}
 {\bf D}_{16}^{\; 3} E_{\mbox{\DSOV{0}{\mathnormal{s}}000}}&=&  \frac{2s(2s-7)+3}{4} {\bf D}_{16} E_{\mbox{\DSOV{0}{\mathnormal{s}}000}} \ ,\quad   {\bf D}_{10}^{\; 3}  E_{\mbox{\DSOV{0}{\mathnormal{s}}000}}= \frac{s(2s-7)+3}{2} {\bf D}_{10}E_{\mbox{\DSOV{0}{\mathnormal{s}}000}}\ , \CR
\Delta  E_{\mbox{\DSOV{0}{\mathnormal{s}}000}}&=&2s(2s-7) E_{\mbox{\DSOV{0}{\mathnormal{s}}000}}\ , 
\eea
such that the regularised Eisenstein series with the normalisation $\frac{4\xi(5) \zeta(6)\xi(8)}{27\xi(2)\xi(4)} \hat{E}{\DSOV{0}{\stfrac{7}{2}}{0}{0}{0}} $ is a homogeneous solution to equation \eqref{EquaF1} that reproduces precisely the anomalous term in \eqref{AnomaLaplace}. Therefore, the `regular' function\footnote{We call this function regular because the absence of anomalous term on the right-hand side of~\eqref{regLap} suggests that it lies in a continuous family of functions that would be  regular at this point.}
\be 
\mathcal{F}^{\rm \scriptscriptstyle R}_\gra{0}{1}   =  \mathcal{F}_\gra{0}{1}  - \frac{4\xi(5) \zeta(6)\xi(8)}{27\xi(2)\xi(4)} \hat{E}_{\DSOV{0}{\stfrac{7}{2}}{0}{0}{0}} \ , 
\ee
defines a particular solution to \eqref{EquaF1} that is a solution to the Laplace equation 
\be 
\label{regLap} 
\Delta \mathcal{F}^{\rm \scriptscriptstyle R}_\gra{0}{1} = - \Scal{ 2\zeta(3) E_{\DSOV{\stfrac{3}{2}}{0}{0}{0}{0}}}^2  \ . 
\ee
By construction, the 3-loop contribution to $ \mathcal{F}_\gra{0}{1} $ is the corresponding constant term of the regularised Eisenstein series  $\hat{E}{\DSOV{0}{\stfrac{7}{2}}{0}{0}{0}} $, as well as the non-analytic term linear in $\phi$ associated to the logarithmic ultra-violet divergence. So instead of computing explicitly the solution to the tensorial differential equation \eqref{EquaF1}, one could have instead used the property that the regularised Eisenstein series  $\hat{E}{\DSOV{0}{\stfrac{7}{2}}{0}{0}{0}} $ is the unique automorphic homogeneous solution to this differential equation, and use the Langlands constant term formula to derive the solution \eqref{SolTeDif}. This computation can easily be generalised to arbitrary dimensions, and this is the approach we shall expand on and follow in the next section to derive the non-analytic components of $\cE_\gra01$ in dimensions four and five. 
 
 \section{Divergent pieces across various dimensions}
\label{sec:Divs}

Our strategy for determining the coefficients in~\eqref{D6R4eqn} outlined at the end of the preceding section can also be stated as follows. We decompose the complete threshold function as
\begin{align}
\mathcal{E}_{\gra01}^{\ord{D}} = \mathcal{F}_{\gra01}^{\ord{D}\, {\rm \scriptscriptstyle R} }+\mathcal{H}_{\gra01}^{\ord{D}}  ,
\end{align} 
where $ \mathcal{F}_{\gra01}^{\ord{D}\, {\rm \scriptscriptstyle R} }$ is a particular (regular) automorphic solution of the non-anomalous inhomogeneous equation 
\be \left( \Delta - \frac{6(14-D)(D-6)}{D-2}\right) \mathcal{F}_{\gra01}^{\ord{D}\, {\rm \scriptscriptstyle R} }  =-\left(\mathcal{E}_{\gra00}^{\ord{D}}\right)^2 \ , \ee
and the tensor equation associated to the adjoint Eisenstein series with $s=\frac{11}{2},\, 6,\, \frac{9}{2}$ and $\frac{7}{2}$ in dimension $D=3$, $4$, $5$ and $6$, respectively \cite{Bossard:2015uga}. It is very important that  we choose  $ \mathcal{F}_{\gra01}^{\ord{D}\, {\rm \scriptscriptstyle R} }$ such that it does not include any three-loop contribution in string perturbation theory. We can always do this because the three-loop contribution is generally the solution to a homogeneous equation. In fact the anomalous contribution is always associated to the three-loop contribution: in six dimensions through the 3-loop ultraviolet divergence, in five dimension through the 2-loop divergence of the 1-loop $R^4$ type form factor, and in four dimensions through the 1-loop divergence of the 2-loop $\nabla^4 R^4$ type form factor. 

From this definition it follows that $\mathcal{H}_{\gra01}^{\ord{D}} $ is a solution to the anomalous (almost) homogeneous equation 
\begin{align}
 \left( \Delta - \frac{6(14-D)(D-6)}{D-2}\right) \mathcal{H}_{\gra01}^{\ord{D}} = \beta_D E_{{\rm Ad},s=\sfrac{(6-D)(1+D)}{4}} \ , 
\end{align}
where $E_{{\rm Ad},s}$ is the adjoint Eisenstein series of $E_{11-D}$ for $D\le 6$ (which is $1$ at $D=6$), and $\beta_D$ is the corresponding numerical coefficient, which will be determined in the sequel. Such a right-hand side must be considered whenever power-counting a priori allows for a form factor or a genuine amplitude logarithmic divergence in supergravity and the occurrence of the adjoint Eisenstein series is fixed by the differential equations of~\cite{Bossard:2015uga}. Supersymmetry Ward identities moreover imply that such a correction can only occur if the two functions satisfy to compatible differential equations and in particular if the Laplace eigenvalues are identical. Investigation gives that this only occurs for $D=6,\, 5,\, 4$. $\mathcal{H}_{\gra01}^{\ord{D}} $ is the sum of two respective solutions to the two tensorial differential equations associated to the two distinct $\nabla^6 R^4$ type invariants \cite{Bossard:2015uga}. Up to cusp forms, the automorphic solution (with appropriate fall off at the boundary of moduli space) to these differential equations is unique, and the T-duality invariant solution is also unique at a given order in string perturbation theory. This implies that the general T-duality invariant solution to the differential equations relevant at 3-loop is necessarily the string theory limit of the fully automorphic solution. We can therefore assume without loss of generality that $\mathcal{H}_{\gra01}^{\ord{D}}$ is the sum of two regularised Eisenstein series solving the  two relevant tensorial differential equations, \ie an adjoint Einsenstein series with  $s=\frac{11}{2},\, 6,\, \frac{9}{2}$ and $\frac{7}{2}$ in dimension 3, 4, 5 and 6, respectively and a series in the fundamental representation associated to the last Dynkin node in the $E_{11-D}$ convention with $s= \frac{14-D}{2}$, which define the same function in three dimensions.

Note that in general the adjoint Eisenstein series admits an expansion in the perturbative string theory limit that is incompatible with string perturbation theory. There is for example always a term that would formally  contribute to a $-\frac{1}{2}$-loop correction. One can easily understand this in type IIB supergravity in ten dimensions, where one can define $\mathcal{F}_{\gra01}^{\ord{10}\, {\rm \scriptscriptstyle R} }$ such that it does not include a 3-loop contribution, in which case 
\be \mathcal{H}_{\gra01}^{\ord{10}} = \frac{4 \zeta(6)\xi(8)}{27 \xi(7)} E_{[4]}(\Omega)  =   \frac{4 \zeta(6)\xi(8)}{27 \xi(7)} \Omega_2^{\, 4} + \frac{4\zeta(6)}{27} \Omega_2^{-3} + \mathcal{O}(e^{-2\pi \Omega_2}) \  ,  \ee
where the first factor is a spurious  $-\frac{1}{2}$-loop contribution. These spurious contributions all occur at negative order, and one understands that they are precisely compensated by the particular solution $\mathcal{F}_{\gra01}^{\ord{D}\, {\rm \scriptscriptstyle R} }$. 

\subsection{Three-loop divergence in $D=6$}

The following (formal) Eisenstein series are the two automorphic homogeneous solutions to the two respective tensorial differential equations \eqref{InhomoD5} and \eqref{HomoD5}
\begin{align}
E_{\DSOV{0}{0}{0}{0}{4}} \quad\textrm{and}\quad E_{\DSOV{0}{\frac72}{0}{0}{0}}\ .
\end{align}
Both of them are singular and need to be regularised.\footnote{In the next section we will confirm by an analysis of their Fourier coefficients that they contribute to the two different $\nabla^6 R^4$ invariants.} A combination that provides a regular limit is
\begin{align}
\label{F3ansatzNew}
\mathcal{H}_{\gra01}^{\ord{6}} =a  
\lim_{\epsilon\to 0} \left(E_{\DSOV{0}{0}{0}{0}{4-\epsilon}} +\frac{\xi(5)}{\xi(2)}E_{\DSOV{0}{\frac72+\epsilon}{0}{0}{0}}
\right),
\end{align}
with $a$ some constant that will be determined in the sequel. The relative coefficient here was chosen to yield a finite limit although this requirement is not forced on us. We will see in the following that it is indeed determined to match the three-loop amplitude threshold functions in ten dimensions. The string perturbation limits of the two series are given by ($\gs = e^\phi$ is the effective string coupling in $D=6$)
\begin{subequations}
\label{pert6}
\begin{align}
\int_{U^1_1} E_{\DSOV{0}{0}{0}{0}{4-\epsilon}} &= \gs^{-4+\epsilon} E_{\DSOIVS{0}{0}{0}{4-\epsilon}} + \gs^{-\epsilon} \frac{\xi(4-2\epsilon)}{\xi(8-2\epsilon)} E_{\DSOIVS{0}{0}{3-\epsilon}{0}}\ ,\\
\label{adjPert}
\int_{U^1_1} E_{\DSOV{0}{\frac72+\epsilon}{0}{0}{0}} &= \gs^{-7-2\epsilon} E_{\DSOIVS{\frac72+\epsilon}{0}{0}{0}} +\gs^{-2} \frac{\xi(5+2\epsilon)}{\xi(7+2\epsilon)}E_{\DSOIVS{0}{3+\epsilon}{0}{0}} \nn\\
&\quad +\gs^{2\epsilon} \frac{\xi(2+2\epsilon)\xi(4+2\epsilon)\xi(7+4\epsilon)}{\xi(5+2\epsilon)\xi(7+2\epsilon)\xi(8+4\epsilon)} E_{\DSOIVS{3+\epsilon}{0}{0}{0}}\ .
\end{align}
\end{subequations}
The poles at $\epsilon\to 0$ are contained in the last terms in both lines (corresponding to three loops after the Weyl rescaling by $\gs^4$). The singular terms are explicitly given by
\begin{subequations}
\begin{align}
E_{\DSOIVS{0}{0}{3-\epsilon}{0}} &= -\frac1{2\epsilon} \cdot \frac{\xi(3)}{\xi(4)\xi(6)} + O(\epsilon^0)\ ,\\
E_{\DSOIVS{3+\epsilon}{0}{0}{0}} &= \frac1{2\epsilon} \cdot \frac{\xi(3)}{\xi(4)\xi(6)} + O(\epsilon^0)\ .
\end{align}
\end{subequations}
Note that these two functions are indeed related by T-duality parity. Demanding T-duality invariance therefore fixes the relative coefficient between the two Eisenstein series, as anticipated in \eqref{F3ansatzNew}. From this one can see that the combination given in~\eqref{F3ansatzNew} is regular. The non-analytic term in $\gs$ can be deduced from this to be
\begin{align}
\label{div}
\mathcal{H}_{\gra01}^{\ord{6}}\to a \frac{\xi(3)}{\xi(6)\xi(8)}  \lim_{\epsilon\to 0} \frac{-\gs^{-\epsilon}+\gs^{2\epsilon}}{2\epsilon}+\ldots = \frac{3}{2}a\frac{\xi(3)}{\xi(6)\xi(8)} \log \gs + \ldots,
\end{align}
where the adjoint series contributes twice the amount of the spinor series (due to the $\gs^{2\epsilon}$ in~\eqref{adjPert}). We note that the perturbative limit of the series $E{\DSOV{0}{\frac72+\epsilon}{0}{0}{0}}$ also contains a term with $\gs^{-7}$ that is formally at loop order $L=-1/2$. This does not matter for determining the logarithmic divergence since this term is cancelled by a corresponding contribution from the particular solution $\mathcal{F}^{\ord{6}\, {\scriptscriptstyle \rm R}}_{\gra01}$, as discussed in the introduction to this section. 

The value of the coefficient $a$ can be fixed by comparison with the ten-dimensional three-loop correction in the decompactification limit ~\cite{Green:2005ba,Gomez:2013sla,Green:2014yxa}. In a first step, one decompactifies the three-loop terms in~\eqref{pert6} to $D=7$ to obtain
\begin{subequations}
\begin{align}
\int_{U^1_4} E_{\DSOIVS{0}{0}{3-\epsilon}{0}} &= v^{3-\epsilon} E_{\DSLIII{0}{0}{3-\epsilon}} + v^\epsilon \frac{\xi(3-2\epsilon)}{\xi(6-2\epsilon)} E_{\DSLIII{2-\epsilon}{0}{0}},\\
\int_{U^1_4} E_{\DSOIVS{3+\epsilon}{0}{0}{0}} &= v^{3+\epsilon} E_{\DSLIII{3+\epsilon}{0}{0}} + v^{-\epsilon} \frac{\xi(3+2\epsilon)}{\xi(6+2\epsilon)} E_{\DSLIII{0}{0}{2+\epsilon}}.
\end{align}
\end{subequations}
Here, $v$ is related to the radius of the decompactifying circle from $D=6$ to $D=7$ measured in seven-dimensional Planck units. Diagrammatically, the expansion above uses the node in the lower right corner of the $D_4$ Dynkin diagram, \ie the vector representation node in the string theory convention. The series above are  still divergent in the second term but we are interested in the first terms that decompactify nicely to give
\begin{align}
\mathcal{H}_{\gra01}^{\ord{6}}\to a \frac{\xi(4)}{\xi(8)} v^3 \left(E_{\DSLIII{0}{0}{3}} + E_{\DSLIII{3}{0}{0}} \right) +\ldots
\end{align}
The combinations that decompactifies correctly to $D=10$ is in our conventions~\cite{Green:2010sp}
\begin{align}
\frac{4\zeta(6)}{27} \left(E_{\DSLIII{0}{0}{3}} + E_{\DSLIII{3}{0}{0}} \right) 
\end{align}
and therefore 
\begin{align}
a = \frac{4\zeta(6)\xi(8)}{27\xi(4)}\ .
\end{align} 
Putting this together with~\eqref{div} means
\begin{align}
\label{logTerm}
\mathcal{H}_{\gra01}^{\ord{6}} \to 5\zeta(3) \log \gs +\ldots
\end{align}
which is indeed consistent with the explicit 3-loop divergence computed in \cite{Bern:2008pv}, as exhibited in \cite{Pioline:2015yea} through the analysis displayed in \cite{Green:2010sp}. This suffices to determine the value of the anomalous term in~\eqref{D6R4eqn} by acting with the $SO(5,5)$ Laplacian on the three-loop terms contained in~\eqref{F3ansatzNew}. 
Acting with the Laplace operator \eqref{PertuLaplace}  on the three-loop terms in $\mathcal{H}^{\ord{6}}_{\gra01}$ proceeds through the introduction of regularised $SO(4,4)$ Eisenstein series by
\begin{subequations}
\begin{align}
E_{\DSOIVS{0}{0}{3-\epsilon}{0}} &= -\frac{1}{2\epsilon} \cdot \frac{\xi(3)}{\xi(4)\xi(6)} + \hat{E}_{\DSOIVS{0}{0}{3}{0}} + O(\epsilon),\\
E_{\DSOIVS{3+\epsilon}{0}{0}{0}} &= \frac{1}{2\epsilon} \cdot \frac{\xi(3)}{\xi(4)\xi(6)} + \hat{E}_{\DSOIVS{3}{0}{0}{0}} + O(\epsilon).
\end{align}
\end{subequations}
Then
\begin{align}
\Delta_{D_4} \hat{E}_{\DSOIVS{0}{0}{3}{0}} 
&= \frac{3\xi(3)}{\xi(4)\xi(6)}.
\end{align}
and the other $SO(4,4)$ series works similarly. Acting with the $SO(5,5)$ Laplacian~\eqref{PertuLaplace} (with $\gs=e^\phi$) thus yields
\begin{align}
\Delta_{D_5} \mathcal{H}^{\ord{6}}_{\gra01} &= \Delta_{D_5} \left[ 5\zeta(3) \log \gs +\frac{4\zeta(6)}{27} \left(\hat{E}_{\DSOIVS{0}{0}{3}{0}}+\hat{E}_{\DSOIVS{3}{0}{0}{0}}\right)\right]\nn\\
&=20\zeta(3) + \frac{4\zeta(6)}{27\xi(4)} \frac{6\xi(3)}{\xi(6)}  \nn\\
&=40 \zeta(3).
\end{align}
This confirms the coefficient in~\eqref{D6R4eqn}. The final function giving rise to the divergent term in $D=6$ is then
\begin{align}
\mathcal{H}_{\gra01}^{\ord{6}} &= \frac{4\zeta(6)\xi(8)}{27\xi(4)}  
\lim_{\epsilon\to 0} \left(E_{\DSOV{0}{0}{0}{0}{4-\epsilon}} +\frac{\xi(5)}{\xi(2)}E_{\DSOV{0}{\frac72+\epsilon}{0}{0}{0}}
\right)\nn\\
&= \frac{4\zeta(6)\xi(8)}{27\xi(4)}  
\left(\hat{E}_{\DSOV{0}{0}{0}{0}{4}} +\frac{\xi(5)}{\xi(2)}\hat{E}_{\DSOV{0}{\frac72}{0}{0}{0}}
\right).
\end{align}

\subsection{Adjoint Eisenstein series of $E_{8(8)}$ and $\nabla^6R^4$ in three dimensions}

Even though the $\nabla^6R^4$ threshold function is regular in three dimensions, it is convenient to start our analysis at $D=3$, because there is a unique $\nabla^6R^4$ supersymmetry invariant \cite{Bossard:2015uga}. The exact threshold function must satisfy the fourth order differential equation
\be 
\label{tensadjE8}
\scal{ \cD \Gamma_{i[jk}{}^r \cD} \scal{ \cD \Gamma_{lpq]r} \cD } \cE_{\gra{0}{1}}^{\ord{3}} = 150 \delta_{i[j} \scal{ \cD \Gamma_{klpq]} \cD}  \cE_{\gra{0}{1}}^{\ord{3}} + \delta_{i[j} \scal{ \cD \Gamma_{klpq]} \cD} \left(\cE_{\gra{0}{0}}^{\ord{3}}\right)^{ 2}\ , 
\ee
consistently with the inhomogeneous equation \cite{Green:2010kv}
 \be
\Delta \cE_{\gra{0}{1}}^{\ord{3}} = - 198  \cE_{\gra{0}{1}}^{\ord{3}} - \left(\cE_{\gra{0}{0}}^{\ord{3}}\right)^{2} \ . 
\ee
The tensorial equation~\eqref{tensadjE8} implies that the wave-front set is associated to the $\tfrac18$-BPS nilpotent orbit corresponding to extremal black holes in four dimensions (see for example \cite{Bossard:2009we}). We compute in Appendix \ref{E8Diff} that the automorphic solution to the  homogeneous equation 
\be 
\scal{ \cD \Gamma_{i[jk}{}^r \cD} \scal{ \cD \Gamma_{lpq]r} \cD } \mathcal{H}^\ord{3}_{\gra{0}{1}}= 150 \delta_{i[j} \scal{ \cD \Gamma_{klpq]} \cD}  \mathcal{H}^\ord{3}_{\gra{0}{1}} \ , 
\ee
is proportional to the  $E_{8(8)}$ Eisenstein series
\begin{align}
E_{\DEVIII{0}{0}{0}{0}{0}{0}{0}{\frac{11}2}}\ .
\end{align}
This is in some sense the ancestor of all three-loop terms in higher dimensions. For this `adjoint' $E_{8(8)}$ series the Laplace eigenvalue is
\begin{align}
\Delta E_{\DEVIII{0}{0}{0}{0}{0}{0}{0}{s}} = 2s(2s-29) E_{\DEVIII{0}{0}{0}{0}{0}{0}{0}{s}}.
\end{align}
The function is regular at $s=\frac{11}2$.

We can perform the decompactification of this function from $D=3$ to $D=6$ by computing the constant term in the parabolic subgroup associated to the $6^{\rm th}$ node of $E_{8(8)}$ with Levi factor $L_6= GL(1)\times Spin(5,5)\times SL(3)$. For arbitrary $s$ the result is
\begin{align}
\label{E8toD5A2}
\int_{U^1_6} E_{\DEVIII{0}{0}{0}{0}{0}{0}{0}{s}} &= v^3 \frac{\xi(2s-11)\xi(2s-12)\xi(2s-13)}{\xi(2s)\xi(2s-5)\xi(2s-9)} E_{\DSOV{0}{s-\frac{11}{2}}{0}{0}{0}}
+ v^{(2s+1)/4} \frac{\xi(2s-3)}{\xi(2s)} E_{\DSOV{0}{0}{0}{0}{s-\frac32}}\nn\\
&\quad+v^{(15-s)/2} \frac{\xi(2s-18)\xi(2s-19)\xi(2s-20)\xi(4s-29)}{\xi(2s)\xi(2s-5)\xi(2s-9)\xi(4s-28)} E_{\DSOV{0}{0}{0}{s-9}{0}}+\ldots,
\end{align}
where we have only listed the $SL(3)$-singlet contributions. The $GL(1)$ parameter $v$ here is related to the volume of the decompactifying three-torus. The last term in~\eqref{E8toD5A2} is subdominant for $s\to \frac{11}2$ and the first term can be  rewritten according to the functional relation
\begin{align}
E_{\DSOV{0}{s}{0}{0}{0}} = \frac{\xi(2s-4)\xi(2s-5)\xi(2s-6)\xi(4s-7)}{\xi(2s)\xi(2s-1)\xi(2s-2)\xi(4s-6)} E_{\DSOV{0}{\frac72-s}{0}{0}{0}}\ .
\end{align}
Then the dominant pieces become
\begin{align}
 v^3 \frac{\xi(2s-15)\xi(2s-16)\xi(2s-17)\xi(4s-29)}{\xi(2s)\xi(2s-5)\xi(2s-9)\xi(4s-28)} E_{\DSOV{0}{9-s}{0}{0}{0}}
+ v^{(2s+1)/4} \frac{\xi(2s-3)}{\xi(2s)} E_{\DSOV{0}{0}{0}{0}{s-\frac32}}
\end{align}
that tend to
\begin{align}
 v^3 \frac{\xi(8)}{\xi(11)}\left(\frac{\xi(5)}{\xi(2)} \hat{E}_{\DSOV{0}{\frac72}{0}{0}{0}}
+   \hat{E}_{\DSOV{0}{0}{0}{0}{4}}\right)
\end{align}
for $s\to\frac{11}2$. This is precisely the combination appearing in the ansatz~\eqref{F3ansatzNew} in six dimensions. From this we conclude that the $D=3$ ancestor of the correct $D=6$ three-loop divergence is given by
\begin{align}
\label{3Dfn}
\mathcal{H}^\ord{3}_{\gra{0}{1}} = a\frac{\xi(11)}{\xi(8)} E_{\DEVIII{0}{0}{0}{0}{0}{0}{0}{\frac{11}2}} =\underbrace{\frac{4\zeta(6)\xi(11)}{27\xi(4)}}_{=:b} E_{\DEVIII{0}{0}{0}{0}{0}{0}{0}{\frac{11}2}}.
\end{align}
Here, we have defined the normalisation constant $b$. We will now use this finite adjoint $E_8$ function to determine the remaining terms in~\eqref{D6R4eqn} and~\eqref{E01pert}.

\subsection{Form factor divergence in $D=5$}

We consider the decompactification limit of the general adjoint $E_{8(8)}$ function from $D=3$ to $D=5$. The Levi subgroup in this case is $L_7=GL(1)\times SL(2)\times E_{6(6)}$. One has
\begin{align}
\label{E8toE6A1}
\int_{U^1_7} E_{\DEVIII{0}{0}{0}{0}{0}{0}{0}{s}} = 
v^4 \frac{\xi(2s-12)\xi(2s-11)}{\xi(2s)\xi(2s-5)} E_{\DEVI{0}{s-\frac92}{0}{0}{0}{0}} 
+v^{(2s+1)/3} \frac{\xi(2s-2)}{\xi(2s)} E_{\DEVI{0}{0}{0}{0}{0}{s-1}}+\ldots,
\end{align}
where we have focussed on the  pieces that are relevant for the discussion. These are in particular singlets under the $SL(2)$ group of the decompactifying two-torus whose size is related to the parameter $v$. The first term in the above expression is clearly divergent for $s\to\frac{11}2$ since $\xi(0)$ diverges and the adjoint $E_{6(6)}$ function is regular at $s=1$. 
We leave the divergence implicit in the second term in~\eqref{E8toE6A1}.
 
We will be interested in the string perturbation limit of the individual terms. For the adjoint $E_{6(6)}$ series one finds from~\eqref{CTmax}
\begin{align}
\int_{U^1_1}  E_{\DEVI{0}{s}{0}{0}{0}{0}}  &= \gs^{-2s} E_{\DSOV{0}{0}{0}{0}{s}} 
+ \gs^{-4} \frac{\xi(2s-4)}{\xi(2s)} E_{\DSOV{0}{s-1}{0}{0}{0}}\nn\\
&\quad + \gs^{2s-11} \frac{\xi(2s-7)\xi(2s-5)\xi(4s-11)}{\xi(2s)\xi(2s-2)\xi(4s-10)} E_{\DSOV{0}{0}{s-\frac32}{0}{0}},
\end{align}
where these are now functions on the symmetric space $SO(5,5)/(SO(5)\times SO(5))$ parametrised by the NS moduli in five dimensions, with the standard $D_5$ labelling and $\gs$ is the five-dimensional string coupling. According to~\eqref{E8toE6A1} we need this expression in the limit $s\to 1+\epsilon$. In this limit the last term disappears and we are left with (recalling the $\xi$ prefactors from~\eqref{E8toE6A1})
\begin{align}
 \frac{\xi(2)\xi(2\epsilon)}{\xi(6)\xi(11)}\left(\gs^{-2-2\epsilon}E_{\DSOV{0}{0}{0}{0}{1}}  + \gs^{-4} \frac{\xi(3)}{\xi(2)} + O(\epsilon)\right).
\end{align}
The string perturbation expansion of the fundamental $E_{6(6)}$ series in~\eqref{E8toE6A1} is
\begin{align}
\int_{U^1_1} E_{\DEVI{0}{0}{0}{0}{0}{s}} &= \gs^{-8(6-s)/3} \frac{\xi(2s-11)\xi(2s-8)}{\xi(2s)\xi(2s-3)} + \gs^{-(15-2s)/3} \frac{\xi(2s-5)}{\xi(2s)} E_{\DSOV{0}{0}{0}{0}{s-\frac32}} \nn\\
&\quad + \gs^s E_{\DSOV{s}{0}{0}{0}{0}},
\end{align}
which exhibits the expected pole when $s\to\frac92$ (which is the right value after taking into account the shift from above). More precisely, for $s=\frac92+\epsilon$ we have the terms (recalling the finite $\xi(9)/\xi(11)$ from~\eqref{E8toE6A1})
\begin{align}
&\quad\quad \gs^{-4+8\epsilon/3}\frac{\xi(3)\xi(1+2\epsilon)}{\xi(6)\xi(11)} + \gs^{-2+2\epsilon/3} \frac{\xi(4)}{\xi(11)} E_{\DSOV{0}{0}{0}{0}{3+\epsilon}} \\
&=\gs^{-4+8\epsilon/3}\frac{\xi(3)\xi(1+2\epsilon)}{\xi(6)\xi(11)} +\gs^{-2+2\epsilon/3} \frac{\xi(2)\xi(1+2\epsilon)}{\xi(6)\xi(11)}  E_{\DSOV{0}{0}{0}{1-\epsilon}{0}},
\end{align}
where we have used a functional equation in the second step.

We collect all the relevant terms from above and find for $s=\frac{11}2+\epsilon$ (ignoring the $v^4$ that only indicates the decompactification)
\begin{align}
&\frac1{2\epsilon}\frac{\xi(2)}{\xi(11)\xi(6)} \left(- \gs^{-2-2\epsilon} E_{\DSOV{0}{0}{0}{0}{1+\epsilon}}-\gs^{-4}\frac{\xi(3)}{\xi(2)}
+  \gs^{-4+8\epsilon/3} \frac{\xi(3)}{\xi(2)} + \gs^{-2+2\epsilon/3}E_{\DSOV{0}{0}{0}{1-\epsilon}{0}}\right)\nn\\
& \to  \frac{4\xi(3)}{3\xi(11)\xi(6)} \log \gs \left(\gs^{-4} +\gs^{-2}\frac{\xi(2)}{\xi(3)}E_{\DSOV{0}{0}{0}{1}{0}}\right).
\end{align}
Here, we have used that $E\DSOV{0}{0}{0}{0}{1}=E\DSOV{0}{0}{0}{1}{0}$ (for this particular value of $s=1$) as required by T-duality.  Multiplying by the normalisation $b$ from~\eqref{3Dfn} one has then as for the logarithmic term in $\gs$:
\begin{align}
\label{5dpert}
\frac{20}{9} ( 2\zeta(3))\log \gs\left(g^{-4} +\gs^{-2}\frac{\xi(2)}{\xi(3)}E_{\DSOV{0}{0}{0}{1}{0}}\right)
=\frac{20}9 \mathcal{E}_{\gra00}^{\ord{5}} \log \gs\  ,
\end{align}
where we have used that the $E_{6(6)}$ Eisenstein series appearing in the $R^4$ corrections has the string perturbation expansion
\begin{align}
\int_{U^1_1} E_{\DEVI{\frac32}{0}{0}{0}{0}{0}} = \gs^{-3} + \gs^{-2} \frac{\xi(2)}{\xi(3)} E_{\DSOV{0}{0}{0}{1}{0}}
\end{align}
and $\mathcal{E}_{\gra00}^{\ord{5}} = 2\zeta(3) E\DEVI{\frac32}{0}{0}{0}{0}{0}$.
The coefficient in~\eqref{5dpert} matches the claimed coefficient in~\eqref{E01pert}.

The next task is to  determine the coefficient in the Laplace equation. This one gets by acting with the Laplacian on the regularised series. To this end we note the following functional relations
\begin{align}
E_{\DEVI{0}{s-\frac92}{0}{0}{0}{0}} = \frac{\xi(2s-19)\xi(2s-17)\xi(2s-16)\xi(4s-29)}{\xi(2s-12)\xi(2s-11)\xi(2s-9)\xi(4s-28)}E_{\DEVI{0}{10-s}{0}{0}{0}{0}}\ .
\end{align}
This absorbs the pole in the first term in~\eqref{E8toE6A1}, leading to
\begin{align}
v^4 \frac{\xi(2s-19)\xi(2s-17)\xi(2s-16)\xi(4s-29)}{\xi(2s)\xi(2s-5)\xi(2s-9)\xi(4s-28)} E_{\DEVI{0}{10-s}{0}{0}{0}{0}} 
+v^{(2s+1)/3} \frac{\xi(2s-2)}{\xi(2s)} E_{\DEVI{0}{0}{0}{0}{0}{s-1}}\ .
\end{align}
Both Eisenstein series are singular and need to be regularised. The Laplace eigenvalues for $s=\frac{11}2+\epsilon$ are
\begin{align}
\Delta E_{\DEVI{0}{\frac92-\epsilon}{0}{0}{0}{0}} &= 2(1+\epsilon)(2\epsilon-9)E_{\DEVI{0}{\frac92-\epsilon}{0}{0}{0}{0}}\ ,\\
\Delta E_{\DEVI{0}{0}{0}{0}{0}{\frac92+\epsilon}} &= \left(\frac{8}3\epsilon^2 +8\epsilon-18\right)E_{\DEVI{0}{0}{0}{0}{0}{\frac92+\epsilon}}\ .
\end{align}
and from the analysis of the poles above we know that we can define the regularised series (denoted with a hat) by
\begin{align}
\frac{\xi(8)\xi(9)}{\xi(2)\xi(11)} E_{\DEVI{0}{\frac92-\epsilon}{0}{0}{0}{0}} &= -\frac1{2\epsilon} \frac{\xi(3)}{\xi(6)\xi(11)} E_{\DEVI{\frac32}{0}{0}{0}{0}{0}}  +\frac{\xi(8)\xi(9)}{\xi(2)\xi(11)}\hat{E}_{\DEVI{0}{\frac92}{0}{0}{0}{0}} +O(\epsilon),\\
\frac{\xi(9)}{\xi(11)}E_{\DEVI{0}{0}{0}{0}{0}{\frac92+\epsilon}} &= \frac{1}{2\epsilon} \frac{\xi(3)}{\xi(6)\xi(11)} E_{\DEVI{\frac32}{0}{0}{0}{0}{0}}  +\frac{\xi(9)}{\xi(11)}\hat{E}_{\DEVI{0}{0}{0}{0}{0}{\frac92}}+O(\epsilon).
\end{align}
Therefore we find the anomalous term in the Laplace equation for $\mathcal{E}_{\gra01}^{\ord{5}}$ to be
\begin{align}
(\Delta+18) \mathcal{E}_{\gra01}^{\ord{5}} &=b \lim_{\epsilon\to 0} \frac{1}{2\epsilon}\left(-2(1+\epsilon)(2\epsilon-9)+\left(\frac{8}3\epsilon^2 +8\epsilon-18\right)
\right)\frac{\xi(3)}{2\zeta(3)\xi(6)\xi(11)}\mathcal{E}_{\gra00}^{\ord{5}}\nn\\
&=b \frac{11\xi(3)}{2\zeta(3)\xi(6)\xi(11)}\mathcal{E}_{\gra00}^{\ord{5}} = \frac{55}3 \mathcal{E}_{\gra00}^{\ord{5}}.
\end{align}
Here, we have used the normalisation $b$ from~\eqref{3Dfn}. This matches the claimed coefficient in~\eqref{D6R4eqn}.

\subsection{Form factor divergence in $D=4$}

We need to consider the decompactification of the adjoint $E_{8(8)}$ series to $D=4$. The Levi subgroup is now $L_8=GL(1)\times E_{7(7)}$. The constant term formula~\eqref{CTmax} leads to a number of terms of which we only display the ones relevant for the derivation of the form factor divergence:
\begin{align}
\label{E8toE7}
\int_{U^1_8} E_{\DEVIII{0}{0}{0}{0}{0}{0}{0}{s}}
&= r^{6} \frac{\xi(2s-17)\xi(2s-19)\xi(2s-22)\xi(4s-29)}{\xi(2s)\xi(2s-5)\xi(2s-9)\xi(4s-28)} E_{\DEVII{\frac{23}2-s}{0}{0}{0}{0}{0}{0}} \nn\\
&\quad+ r^{(2s+1)/2} \frac{\xi(2s-1)}{\xi(2s)} E_{\DEVII{0}{0}{0}{0}{0}{0}{s-\frac12}} +\ldots.
\end{align}
Both coefficients are regular but the two Eisenstein series diverge at $s=\frac{11}2$. The parameter $r$ here is related to size of the decompactifying circle. 

Next, we require the string perturbation limit of the two Eisenstein series on $E_{7(7)}/(SU(8)/\mathds{Z}_2) $. For the adjoint $E_{7(7)}$ series one has
\begin{align}
\int_{U^1_1} E_{\DEVII{\frac{23}2-s}{0}{0}{0}{0}{0}{0}}  &= \gs^{-4(s-3)}\frac{\xi(2s-6)\xi(2s-9)\xi(2s-11)\xi(4s-28)}{\xi (2s-17) \xi (2s-19) \xi (2s-22) \xi(4s-29)}\nn\\
&\quad+ \gs^{-8} \frac{\xi (2s-11) \xi (2s-14)}{\xi (2s-19) \xi (2s-22)}E_{\DSOVI{15-2s}{s-\frac{11}2}{0}{0}{0}{0}}\\
&\quad+\gs^{5-2s}\frac{\xi (2s-7) \xi (2s-9) \xi (2s-11) \xi (4s-28)}{\xi (2s-17) \xi (2s-19) \xi (2s-22) \xi
   (4s-29)} E_{\DSOVI{0}{0}{0}{0}{0}{s-\frac{7}2}} +\ldots.\nn
\end{align}
We have chosen a representative that brings out the divergence at $s=\frac{11}2+\epsilon$ explicitly through the factor $\xi(2s-11)= \xi(2\epsilon) = -\frac{1}{2\epsilon} + \ldots$. For $s=\frac{11}2+\epsilon$ one then has (after reinstating the prefactor from~\eqref{E8toE7} and not exhibiting the $r$ dependence)
\begin{align}
\label{D4div1}
&\quad \frac{\xi(2s-17)\xi(2s-19)\xi(2s-22)\xi(4s-29)}{\xi(2s)\xi(2s-5)\xi(2s-9)\xi(4s-28)}\int_{U^1_1} E_{\DEVII{\frac{23}2-s}{0}{0}{0}{0}{0}{0}}  \\
& \to -\frac{1}{2\epsilon}\frac{\xi(5)}{\xi(6)\xi(11)}  E_{\DEVII{\frac52}{0}{0}{0}{0}{0}{0}} 
+\frac{\xi(5)}{\xi(6)\xi(11)} \log \gs \left( 2\gs^{-10}+ \gs^{-6}\frac{\xi(4)}{\xi(5)} E_{\DSOVI{0}{0}{0}{0}{0}{2}}\right) +\ldots,\nn
\end{align}
where we have used the string perturbative expansion 
\begin{align}
\int_{U^1_1} E_{\DEVII{\frac52}{0}{0}{0}{0}{0}{0}} = \gs^{-10} + \gs^{-8}  \frac{\xi(4)\xi(8)}{\xi(2)\xi(5)} E_{\DSOVI{4}{0}{0}{0}{0}{0}} + \gs^{-6} \frac{\xi(4)}{\xi(5)}E_{\DSOVI{0}{0}{0}{0}{0}{2}}.
\end{align}

For the fundamental $E_{7(7)}$ series one finds the perturbative string expansion
\begin{align}
\label{E7toD62}
\int_{U^1_1} E_{\DEVII{0}{0}{0}{0}{0}{0}{s-\frac12}} &= \gs^{1-2s}  \frac{\xi(2s-6)\xi(2s-10)}{\xi(2s-1)\xi(2s-5)} E_{\DSOVI{\frac{11}2-s}{0}{0}{0}{0}{0}}\nn\\
&\quad+\gs^{2s-19} \frac{\xi(2s-10)\xi(2s-14)\xi(2s-18)}{\xi(2s-1)\xi(2s-5)\xi(2s-19)}E_{\DSOVI{\frac{19}{2}-s}{0}{0}{0}{0}{0}} \nn\\
&\quad+\gs^{-6} \frac{\xi(2s-10)\xi(2s-12)\xi(2s-14)}{\xi(2s-1)\xi(2s-5)\xi(2s-9)} E_{\DSOVI{0}{0}{0}{0}{\frac{15}2-s}{0}}.
\end{align}
This expansion is exact and we have again chosen a form that brings out the divergence explicitly through the prefactor $\xi(2s-10)=\xi(2\epsilon)=+\frac{1}{2\epsilon}+\ldots$ for $s=\frac{11}2+\epsilon$. Reinstating the prefactor from~\eqref{E8toE7} one then obtains for the $\frac{1}{\epsilon}$ and $\log \gs$ terms 
\begin{align}
\label{D4div2}
&\quad \frac{\xi(2s-1)}{\xi(2s)} \int_{U^1_1} E_{\DEVII{0}{0}{0}{0}{0}{0}{s-\frac12}} \\
&\to \frac1{2\epsilon} \frac{\xi(5)}{\xi(6)\xi(11)} E_{\DEVII{\frac52}{0}{0}{0}{0}{0}{0}}
 +\frac{\xi(5)}{\xi(6)\xi(11)}  \log \gs \left(-\gs^{-10} + \gs^{-8} \frac{\xi(4)\xi(8)}{\xi(2)\xi(5)} E_{\DSOVI{4}{0}{0}{0}{0}{0}} \right).\nn
\end{align}

Putting~\eqref{D4div1} and~\eqref{D4div2} together we see that the divergence indeed cancels out and the $\log  \gs$ terms are
\begin{align}
&\quad \frac{\xi(5)}{\xi(6)\xi(11)}  \log \gs \left( \gs^{-10} +\gs^{-8} \frac{\xi(4)\xi(8)}{\xi(2)\xi(5)} E_{\DSOVI{4}{0}{0}{0}{0}{0}} +\gs^{-6}\frac{\xi(4)}{\xi(5)} E_{\DSOVI{0}{0}{0}{0}{0}{2}}
\right) \nn\\
&= \frac{\xi(5)}{\xi(6)\xi(11)}  \log \gs \, E_{\DEVII{\frac52}{0}{0}{0}{0}{0}{0}}.
\end{align}
The reassembly of the full adjoint $E_{7(7)}$ series was only done at the level of the string perturbative terms but will hold fully. 
Multiplying in the normalisation $b$ from~\eqref{3Dfn} the final result is\footnote{Recall that $\mathcal{E}_{\gra10}^{\ord{4}}=\zeta(5) E\DEVII{\frac52}{0}{0}{0}{0}{0}{0}$ without a factor of $2$.}
\begin{align}
\frac{\xi(5)}{\xi(6)\xi(11)} b \log \gs\,  E_{\DEVII{\frac52}{0}{0}{0}{0}{0}{0}}
= \frac{5}{\pi} \zeta(5) \log \gs  \, E_{\DEVII{\frac52}{0}{0}{0}{0}{0}{0}}
= \frac{5}{\pi} \log \gs \, \mathcal{E}_{\gra10}^{\ord{4}}.
\end{align}
This confirms the relevant term in~\eqref{E01pert}.

We now turn to the computation of the anomalous term in the Laplace equation. We define regularisations (denoted by hats) of the series appearing in~\eqref{E8toE7} in accordance with the pole analysis at $s=\frac{11}2+\epsilon$ by
\begin{align}
\frac{\xi(8)\xi(9)\xi(12)}{\xi(2)\xi(6)\xi(11)} E_{\DEVII{6-\epsilon}{0}{0}{0}{0}{0}{0}} &= -\frac{1}{2\epsilon} \frac{\xi(5)}{\xi(6)\xi(11)} E_{\DEVII{\frac52}{0}{0}{0}{0}{0}{0}} + \frac{\xi(8)\xi(9)\xi(12)}{\xi(2)\xi(6)\xi(11)} \hat{E}_{\DEVII{6}{0}{0}{0}{0}{0}{0}} + O(\epsilon),\\
\frac{\xi(10)}{\xi(11)} E_{\DEVII{0}{0}{0}{0}{0}{0}{5+\epsilon}} &= \frac{1}{2\epsilon} \frac{\xi(5)}{\xi(6)\xi(11)} E_{\DEVII{\frac52}{0}{0}{0}{0}{0}{0}} +\frac{\xi(10)}{\xi(11)} \hat{E}_{\DEVII{0}{0}{0}{0}{0}{0}{5}} + O(\epsilon).
\end{align}
Here we have kept the prefactors as they appear in the combination~\eqref{E8toE7}. Using now
\begin{align}
\Delta  E_{\DEVII{6-\epsilon}{0}{0}{0}{0}{0}{0}} &= \left(4\epsilon^2-14\epsilon-60\right)  E_{\DEVII{6-\epsilon}{0}{0}{0}{0}{0}{0}},\\
\Delta  E_{\DEVII{0}{0}{0}{0}{0}{0}{5+\epsilon}} &= \left(3\epsilon^2+3\epsilon-60\right) E_{\DEVII{0}{0}{0}{0}{0}{0}{5+\epsilon}},
\end{align}
one deduces
\begin{align}
\frac{\xi(8)\xi(9)\xi(12)}{\xi(2)\xi(6)\xi(11)} \left(\Delta+60\right) \hat{E}_{\DEVII{6}{0}{0}{0}{0}{0}{0}}  &=  \frac{7\xi(5)}{\xi(6)\xi(11)} E_{\DEVII{\frac52}{0}{0}{0}{0}{0}{0}},\\
\frac{\xi(10)}{\xi(11)} \left(\Delta+60\right) \hat{E}_{\DEVII{0}{0}{0}{0}{0}{0}{5}} &=  \frac{3\xi(5)}{2\xi(6)\xi(11)} E_{\DEVII{\frac52}{0}{0}{0}{0}{0}{0}}.
\end{align}
The anomalous term in the Laplace equation of $\mathcal{E}_{\gra01}^{\ord{4}}$ is therefore
\begin{align}
 \left(\Delta+60\right) \mathcal{E}_{\gra01}^{\ord{4}} = b\left(7+\frac32\right)   \frac{\xi(5)}{\xi(6)\xi(11)} E_{\DEVII{\frac52}{0}{0}{0}{0}{0}{0}} = \frac{85}{2\pi} \mathcal{E}_{\gra10}^{\ord{4}}\ ,
\end{align}
thus confirming~\eqref{D6R4eqn}.

\section{Fourier expansions and Whittaker vectors for $SO(5,5)$}
\label{sec:Fourier}

The Fourier coefficients of the functions $\mathcal{E}_{\gra{p}{q}}^{\ord{D}}$ are constrained by supersymmetry~\cite{Pioline:2010kb,Green:2011vz,Bossard:2014lra}. This can be rephrased in terms of nilpotent orbits~\cite{Green:2011vz} and constraints on degenerate Whittaker vectors~\cite{Fleig:2013psa,Gustafsson:2014iva}. In this section we will investigate these issues for the case of $D=6$ where the Cremmer--Julia group is $SO(5,5)$.

\subsection{Fourier modes}

Given a U-duality invariant function $f(\cosel_D)$ on $G/K$, as for example $\mathcal{E}_{\gra{p}{q}}^{\ord{D}}$, one defines the (abelian) Fourier coefficients in some (abelian) unipotent subgroup $U\subset G$ by
\begin{align}
\label{abFC}
F_{\vec{m}_U} (\cosel_D) = \int_{U^1} f(u \cosel_D) \overline{\psi_{\vec{m}_U} (u)} du.
\end{align}
Here we have used the following notation.
\begin{itemize}
\item $U^1=U(\ints)\backslash U$ where $U(\ints)=U\cap G(\ints)$ denotes the U-duality shifts contained in the chosen unipotent $U$. In an appropriate normalisation the integral is nothing but an integral over $[0,1]^{\dim (U)}$.
\item The unipotent subgroup $U$ corresponds to a choice of subset of moduli in $D$ dimension. All these moduli must be axionic, \ie, have shift symmetries, and their shift symmetries must close on $U$. If $U$ is abelian then all shifts commute but more general cases are possible, for example in the presence of NS$5$-branes~\cite{Becker:1995kb,Pioline:2009qt}. 
\item As for abelian $U$ all shifts commute we can diagonalise them simultaneously and denote the corresponding eigenvalues by $\vec{m}_U\in \ints^{\dim (U)}$. Similarly, we can parametrise any element $u\in U$ by $u=\exp( \vec{\chi}_U \cdot \vec{t}_U)$ where $\vec{t}_U$ denotes a basis of appropriately normalised generators of the unipotent group $U$. The character $\psi_{\vec{m}_U}(u)$ appearing in the formula above then is nothing but
\begin{align}
\psi_{\vec{m}_U}(u) = \exp\left( 2\pi i \vec{m}_U \cdot \vec{\chi}_U \right), \label{PsiFourier} 
\end{align}
\ie a collection of phases associated with the axionic moduli $\vec{\chi}_U$ that are physically defined modulo an integral shift. 
\item If $U$ is non-abelian, then only the abelianised part is captured by the integral~\eqref{abFC}. When $U$ is non-abelian the derived series of $U$ defined by $U^{(k)}=[U^{(k-1)},U^{(k-1)} ]$ (with $U^{(0)}=U$) is non-trivial for $0\leq k \leq n$ for some $n>0$. One can then define similar non-abelian Fourier coefficients for each of the $U^{(k)}$ for $k>0$. 
\end{itemize}
The integers $\vec{m}_U$ correspond physically to instanton charges and the Fourier coefficients arrange themselves on orbits of the action on $U$  of the (reductive) Levi subgroup $L$ associated with the unipotent $U$. The case $\vec{m}_U=0$ corresponds to $\psi_{\vec{m}_U}=1$ and therefore the zero mode Fourier integral reduces to the constant term integral~\eqref{CTmax} discussed in section~\ref{sec:Eisenstein}. 

\subsection{Fourier modes of $SO(5,5)$ series}
\label{sec:FourierD5}

We consider the Fourier modes in an expansion corresponding to decompactification to type IIB. At the level of Lie algebras this means that we are looking at the decomposition of $\mf{so}(5,5)$ under $\mathrm{Lie}\,(L_2)=\mf{gl}(1)\oplus \mf{sl}(2)\oplus \mf{sl}(4)$ where $\mf{sl}(4)\cong \mf{so}(3,3)$ describes the decompactifying four-torus, $\mf{gl}(1)$ is related to its volume and $\mf{sl}(2)$ is the type IIB S-duality. The relevant decomposition is
\begin{align}
 \label{AdjointD5}
\mathfrak{so}(5,5) \cong {\bf1}^\ord{-2} \oplus ( {\bf 2},{\bf 6})^\ord{-1} \oplus \scal{ \mathfrak{gl}(1) \oplus \mathfrak{sl}(2)\oplus \mathfrak{sl}(4)}^\ord{0}  \oplus \underbrace{( {\bf 2}, {\bf 6})^\ord{1} \oplus{\bf1}^\ord{2}}_{\mathrm{Lie}\,(U_2)}
\end{align}
The superscripts in this equation denote the $\mf{gl}(1)$ weight and the representations of $\mf{sl}(2)\oplus\mf{sl}(4)$ are labelled by their dimension. As indicated, the Lie algebra of the non-abelian unipotent subgroup $U_2$ in this expansion consists of two pieces occurring at weights $+1$ and $+2$ with respect to $\mf{gl}(1)$. The associated Heisenberg algebra is realised on the symmetric space $SO(5,5)/(SO(5)\times SO(5))$ through the Killing vectors 
\be 
\kappa^{ij}_\alpha = \frac{\partial\, }{\partial a^\alpha_{ij}} - \frac{1}{4}\varepsilon_{\alpha\beta} \varepsilon^{ijkl} a^\beta_{kl}\frac{\partial\, }{\partial b} \ , \qquad k_5 = \frac{\partial\, }{\partial b} \ , 
\ee
satisfying
\be 
[ \kappa^{ij}_\alpha , \kappa^{kl} _\beta ] = \frac{1}{2}\varepsilon_{\alpha\beta} \varepsilon^{ijkl} k_5 \ . 
\ee
The indices $i,j=1,\ldots,4$ here are fundamental $\mathfrak{sl}(4)$ indices and an antisymmetric pair $[ij]$ corresponds to the vector of $\mathfrak{so}(3,3)\cong \mathfrak{sl}(4)$. The index $\alpha=1,2$ is a fundamental $\mathfrak{sl}(2)$ index. An abelian Fourier mode defined in~\eqref{abFC} here is labelled by an $SL(2,\ints)$ doublet $Q=(p,q)=\vec{m}_U$ of $SO(3,3)$ vectors. The non-abelian Fourier mode associated with the weight $+2$ singlet corresponds to a single integer $N$ and we will write the most general Fourier mode as $F_{Q,N}(\cosel_6)$. Note that this notation is not uniquely defined because not all $\kappa^{ij}_\alpha$ can be diagonalised simultaneously when $N\neq 0$. A related discussion can be found in \cite{Pioline:2009qt}.  For a purely abelian Fourier mode $F_{Q,0}$, the action of the generators is $k_5 F_{Q,0}(\cosel_6) =0$ and $\kappa_\alpha^{ij} F_{Q,0}(\cosel_6) =2\pi i  Q_\alpha^{ij} F_{Q,0} (\cosel_6)$. A non-abelian Fourier coefficient $F_{Q,N}$ satisfies in general  $k_5 F_{Q,N} (\cosel_6)  =2\pi i N F_{Q,N} (\cosel_6)$. 
In physical terms, the Fourier modes $F_{Q,N}$ correspond  to fundamental and D1 strings wrapping the $T^4$ and define $Q_\alpha^{ij}$ whereas the additional single integer $N$ gives the number of Euclidean D3 branes wrapping $T^4$. 

The space of vector doublet charges $(p,q)\in \ints^{12}$ stratifies under the action of $SL(2,\mathds{R})\times SO(3,3)$ according to the number of linearly independent vectors spanned by $p$ and $q$ and the sign of their norm. This stratification is isomorphic to the stratification of real nilpotent orbits of $SO(5,5)$ of dimension smaller than $26$, and is displayed in figure~\ref{ClosureDiagFourier}, where the number of $+,\, -,\, 0$ gives the number of linearly independent vectors whose norm is respectively positive, negative or null.\footnote{In terms of signed partitions of $10$ parametrising the nilpotent orbits of $SO(5,5)$, a $+$ and a $-$ correspond respectively to a $3$-box line with respectively $+-+$ and $-+-$, whereas a $0$ corresponds to a doublet of 2-box lines, the remaining being understood as being $1$-box lines with the appropriate signs to neutralise the partition, see \eg \cite{Bossard:2009we}.}
 \def\xshift{5}
  \def\xmin{1}
 \def\ymin{-2}
\begin{figure}[htbp]
\begin{center}
 \begin{tikzpicture}
  \draw (\xmin,\ymin) node{$\{0\}$};
 \draw (\xmin,\ymin+1) node{$[0]$};
 \draw (\xmin+1,\ymin +2) node{$[+]$};
  \draw (\xmin-1,\ymin +2)  node{$[-]$};
  \draw (\xmin,\ymin +3)  node{$[0,0]$};
  \draw (\xmin+0.9,\ymin +4)  node{$[+,0]$};
  \draw (\xmin -0.9,\ymin +4)  node{$[-,0]$};
   \draw (\xmin +1.4,\ymin +5)  node{$[+,+]$};
    \draw (\xmin - 1.4 ,\ymin +5) node{$[-,-]$};
     \draw (\xmin  ,\ymin +5) node{$[+,-]$};  
       \draw[-,draw=black,very thick](\xmin,\ymin+0.7) -- (\xmin,\ymin +0.3);
  \draw[-,draw=black,very thick](\xmin+0.2,\ymin+1.3) -- (\xmin+0.8,\ymin + 1.7 );
  \draw[-,draw=black,very thick](\xmin-0.2,\ymin+1.3) -- (\xmin-0.8,\ymin + 1.7);
  \draw[-,draw=black,very thick](\xmin,\ymin+1.3) -- (\xmin,\ymin + 2.7 );
    \draw[-,draw=black,very thick](\xmin+1,\ymin+3.7) -- (\xmin+1,\ymin + 2.3 );
      \draw[-,draw=black,very thick](\xmin+0.8,\ymin+3.7) -- (\xmin+0.1,\ymin + 3.3 );
        \draw[-,draw=black,very thick](\xmin+1.1,\ymin+4.3) -- (\xmin+1.3,\ymin + 4.7 );
          \draw[-,draw=black,very thick](\xmin+0.7,\ymin+4.3) -- (\xmin+0.2,\ymin + 4.7);
   \draw[-,draw=black,very thick](\xmin-1,\ymin+3.7) -- (\xmin-1,\ymin + 2.3 );
      \draw[-,draw=black,very thick](\xmin-0.8,\ymin+3.7) -- (\xmin-0.1,\ymin + 3.3 );
            \draw[-,draw=black,very thick](\xmin-1.1,\ymin+4.3) -- (\xmin-1.3,\ymin + 4.7 );
          \draw[-,draw=black,very thick](\xmin-0.7,\ymin+4.3) -- (\xmin-0.2,\ymin + 4.7);
                     \draw (\xmin-3,\ymin) node{$$};
\draw (\xmin-3,\ymin+1) node{$A_1$};
 \draw (\xmin-3,\ymin +2) node{$(2A_1)'$};
  \draw (\xmin-3,\ymin +3)  node{$(2A_1)''$};
  \draw (\xmin-3,\ymin +4)  node{$3A_1$};
   \draw (\xmin -3,\ymin +5)  node{$A_2$};
   
     \draw (\xmin+\xshift,\ymin) node{$(1,1)$};
 \draw (\xmin+\xshift,\ymin+1) node{$(\frac{1}{2},\frac{1}{2})$};
 \draw (\xmin+\xshift+1,\ymin +2) node{$(\frac{1}{2},0)$};
  \draw (\xmin+\xshift-1,\ymin +2)  node{$(0,\frac{1}{2})$};
  \draw (\xmin+\xshift,\ymin +3)  node{$(\frac{1}{4},\frac{1}{4})$};
  \draw (\xmin+\xshift+0.9,\ymin +4)  node{$(\frac{1}{4},0)$};
  \draw (\xmin+\xshift -0.9,\ymin +4)  node{$(0,\frac{1}{4})$};
   \draw (\xmin+\xshift +1.4,\ymin +5)  node{$(\frac{1}{4},0)$};
    \draw (\xmin+\xshift - 1.4 ,\ymin +5) node{$(0,\frac{1}{4})$};
     \draw (\xmin+\xshift  ,\ymin +5) node{$(0,0)$};  
       \draw[-,draw=black,very thick](\xmin+\xshift,\ymin+0.7) -- (\xmin+\xshift,\ymin +0.3);
  \draw[-,draw=black,very thick](\xmin+\xshift+0.2,\ymin+1.3) -- (\xmin+\xshift+0.8,\ymin + 1.7 );
  \draw[-,draw=black,very thick](\xmin+\xshift-0.2,\ymin+1.3) -- (\xmin+\xshift-0.8,\ymin + 1.7);
  \draw[-,draw=black,very thick](\xmin+\xshift,\ymin+1.3) -- (\xmin+\xshift,\ymin + 2.7 );
    \draw[-,draw=black,very thick](\xmin+\xshift+1,\ymin+3.7) -- (\xmin+\xshift+1,\ymin + 2.3 );
      \draw[-,draw=black,very thick](\xmin+\xshift+0.8,\ymin+3.7) -- (\xmin+\xshift+0.1,\ymin + 3.3 );
        \draw[-,draw=black,very thick](\xmin+\xshift+1.1,\ymin+4.3) -- (\xmin+\xshift+1.3,\ymin + 4.7 );
          \draw[-,draw=black,very thick](\xmin+\xshift+0.7,\ymin+4.3) -- (\xmin+\xshift+0.2,\ymin + 4.7);
   \draw[-,draw=black,very thick](\xmin+\xshift-1,\ymin+3.7) -- (\xmin+\xshift-1,\ymin + 2.3 );
      \draw[-,draw=black,very thick](\xmin+\xshift-0.8,\ymin+3.7) -- (\xmin+\xshift-0.1,\ymin + 3.3 );
            \draw[-,draw=black,very thick](\xmin+\xshift-1.1,\ymin+4.3) -- (\xmin+\xshift-1.3,\ymin + 4.7 );
          \draw[-,draw=black,very thick](\xmin+\xshift-0.7,\ymin+4.3) -- (\xmin+\xshift-0.2,\ymin + 4.7);
   \end{tikzpicture}
\end{center}
\caption{\small Closure diagram of the real nilpotent orbits  of $SO(5,5)$ of  dimension smaller than 26. The second copy exhibits the fraction of supersymmetry charges of the two chiralities preserved by the corresponding instanton. The total fraction of supercharges that are preserved is given by half the sum of the two chiral pieces.}
\label{ClosureDiagFourier}
\end{figure}
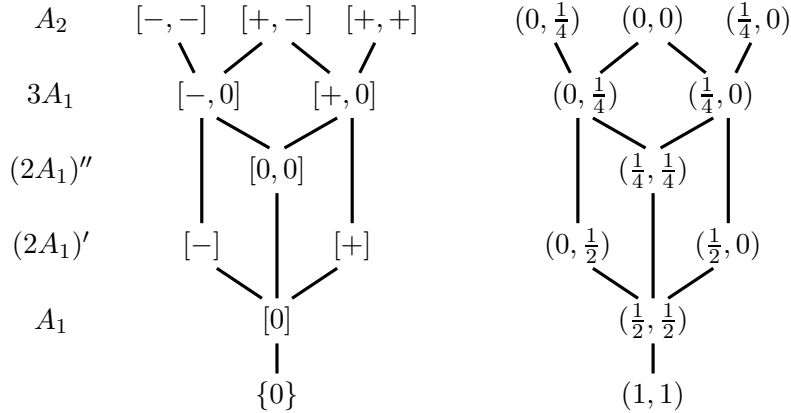

The algebraic constraints defining this stratification are displayed in table~\ref{tab:D5orbs}, where we have also indicated the nilpotent $SO(10,\cx)$ orbits that are intersected by the orbits of the $SO(3,3)\times SL(2)$ action on $\reals^{12}$ and have labelled them both  by their Bala--Carter type and their weighted Dynkin diagram~\cite{BalaCarterI,BalaCarterII,CollingwoodMcGovern}. The qualifier `special' or `not special' refers to Lusztig's property of nilpotent orbits~\cite{Lusztig} and is related to whether they can arise as wave-front sets of automorphic representations~\cite{BarbaschVogan,Jiang:2014}. The information of this table can partially be extracted from~\cite[Sec. 5.5.2]{MillerSahi}. The notation here is such that $|p|^2=p\cdot p$ denotes the $SO(3,3)$-norm of the vector $p$ and $(p \cdot q)^2$ denotes the square of the $SO(3,3)$ inner product. The wedge $p\wedge q$ is a tensorial object, namely the outer product of the two vectors. 
\begin{table}[t!]
\begin{center}
\begin{tabular}{c|c|c|c|c}
Condition on $(p,q)\in \ints^{12}$& `Dimension' of  & \multicolumn{3}{c}{part of nilpotent $SO(10,\cx)$ orbit $\mathcal{O}$}\\
 &subspace & Bala--Carter label & weighted Dynkin&  $\dim_\cx\mathcal{O}$\\
&&of $\mathcal{O}$ &diagram of $\mathcal{O}$ &\\\hline
$|p|^2|q|^2 - (p\cdot q)^2 \neq 0 $ & ${12}$ & $A_2$ (special) &$\DSOV{0}{2}{0}{0}{0}$ & $26$\\
$|p|^2|q|^2 - (p\cdot q)^2 =0 $ & $11 $ & $3A_1$ (not special)&$\DSOV{1}{0}{1}{0}{0}$ & $24$\\
$|p|^2=|q|^2 = (p\cdot q)^2=0 $ & $9$ & $(2A_1)''$ (special)&$\DSOV{0}{0}{0}{1}{1}$ & $20$\\
$p\wedge q =0 $ & $7$ & $(2A_1)'$ (special)&$\DSOV{2}{0}{0}{0}{0}$ & $16$\\
$|p|^2=|q|^2 =p\wedge q = 0 $ & $6$ & $A_1$ (special) &$\DSOV{0}{1}{0}{0}{0}$ & $14$
\end{tabular}
\caption{\label{tab:D5orbs}\small Orbits of $SO(3,3,\ints)\times SL(2,\ints)$ acting on the twelve charges $(p,q)\in \ints^{12}$ describing the (abelian) Fourier coefficients associated to $P_2$. The intersection with the complex nilpotent orbits of $SO(5,5,\cx)\cong SO(10,\cx)$ is also given.}
\end{center}
\end{table}

The non-trivial (abelian) Fourier coefficients must fall into the topological closure of one of the above classes. They correspond to space-time instantons that are defined in the background of supersymmetric solutions in supergravity. These solutions are defined in the Euclidean signature as solutions to the pseudo-Riemannian non-linear sigma model over $G/K^*$ (for a non-compact real form $K^*$ of $K$)  satisfying the energy momentum tensor constraint 
\be 
\langle P_\mu(\Phi) ,P_\nu(\Phi)  \rangle d\Phi^\mu \otimes d\Phi^\nu  = 0  \ .  
\ee
Such a spherically symmetric solution can be obtained for any representative ${\bf p}$ of the nilpotent orbit restricted to the coset component $\mathfrak{g}\ominus\mathfrak{k}^*$, as~\cite{Bossard:2009at} 
\be 
\cosel(\Phi) = \exp\Scal{ -\frac{\bf p}{r^4}} \ ,  
\ee
and the corresponding solution preserves a given amount of supercharges depending of the complex $K(\mathds{C})$ orbit of ${\bf p}$, as is displayed in figure \ref{ClosureDiagFourier}. The smaller the orbit is the stronger the constraints from supersymmetry are, and figure \ref{ClosureDiagFourier} encompasses all the BPS orbits. The differential constraints satisfied by the BPS protected threshold functions are themselves associated to harmonic superspace constructions of the associated linearised supersymmetry invariant as superspace integrals of BPS protected integrands preserving the same amount of supersymmetry, \ie 1/2 BPS for $R^4$, 1/4 for $\nabla^4 R^4$ and 1/8 BPS for $\nabla^6 R^4$ \cite{Bossard:2014lra}. 

For a generic adjoint Eisenstein series
\begin{align}
E_{\mbox{\DSOV{0}{s}000}}
\end{align}
the Fourier coefficients are restricted to the BPS orbits displayed in figure \ref{ClosureDiagFourier}, and generically cover all of them non-trivially, but for special values of $s$ for which they are further restricted. The constraints on the Fourier modes follow from the tensorial differential equations the corresponding function satisfy \cite{Bossard:2014lra,Bossard:2015uga}. One can  check explicitly that these differential equations are satisfied by the generating character of the series, but one must be careful when the Eisenstein series is not absolutely convergent. Indeed we will see in the next subsection, that although the adjoint Eisenstein series at $s=\frac{3}{2}$ is generated by a character satisfying the quartic constraint associated to the dimension $24$ nilpotent orbit of type $3A_1$ in  table~\ref{tab:D5orbs}, its Fourier modes do not satisfy the corresponding constraint $|p|^2 |q|^2 - (p,q)^2 =  0$. This nilpotent orbit is indeed not special, and so one understands this property from the analysis of \cite{BarbaschVogan,Jiang:2014}, that states that the maximal orbits in a wave-front set of an automorphic function are special. There are only two non-special nilpotent orbits of $SO(5,5,\mathds{C})$, as we display in figure \ref{ClosureDiagComplex}. 
 \def\xshift{- 1}
  \def\xmin{1}
 \def\ymin{0}
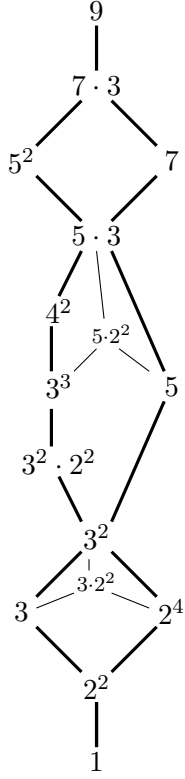
\begin{figure}[t!]
\begin{center}
 \begin{tikzpicture}[rotate=0]
  \draw (\xmin,\ymin) node{$1$};
  \draw (\xmin,\ymin+1) node{$2^2$};
  \draw (\xmin+1,\ymin+2) node{$2^4$};
  \draw (\xmin-1,\ymin+2) node{$3$};
    \draw (\xmin,\ymin+2.4) node{${\scriptstyle 3\cdot 2^2}$};
  \draw (\xmin,\ymin+3) node{$3^2$};
  \draw (\xmin-0.5,\ymin+4) node{$3^2\cdot2^2$};
  \draw (\xmin-0.5,\ymin+5) node{$3^3$};
   \draw (\xmin+1,\ymin+5) node{$5$};
  \draw (\xmin-0.5,\ymin+6) node{$4^2$};
    \draw (\xmin+0.2,\ymin+5.7) node{${\scriptstyle 5\cdot 2^2}$};
\draw (\xmin,\ymin+7) node{$5\cdot3$};
 \draw (\xmin+1,\ymin+8) node{$7$};
  \draw (\xmin-1,\ymin+8) node{$5^2$};
    \draw (\xmin,\ymin+9) node{$7\cdot3$};
    \draw (\xmin,\ymin+10) node{$9$};
  \draw[-,draw=black,very thick](\xmin,\ymin+0.2) -- (\xmin,\ymin 0.8 );
    \draw[-,draw=black,very thick](\xmin+0.2,\ymin+1.2) -- (\xmin+0.8,\ymin 1.8 );
    \draw[-,draw=black,very thick](\xmin-0.2,\ymin+1.2) -- (\xmin-0.8,\ymin 1.8 );
 \draw[-,draw=black,very thick](\xmin+0.8,\ymin+2.2) -- (\xmin+0.2,\ymin 2.8 );
    \draw[-,draw=black,very thick](\xmin-0.8,\ymin+2.2) -- (\xmin-0.2,\ymin 2.8 );
    \draw[-,draw=black,very thick](\xmin-0.2,\ymin+3.2) -- (\xmin-0.5,\ymin 3.8 );
    \draw[-,draw=black,very thick](\xmin-0.6,\ymin+4.2) -- (\xmin-0.6,\ymin 4.7 );
    \draw[-,draw=black,very thick](\xmin-0.6,\ymin+5.2) -- (\xmin-0.6,\ymin 5.8 );
    \draw[-,draw=black,very thick](\xmin-0.2,\ymin+6.8) -- (\xmin-0.5,\ymin 6.2 );
    \draw[-,draw=black,very thick](\xmin+0.2,\ymin+3.2) -- (\xmin+0.9,\ymin 4.8 );
    \draw[-,draw=black,very thick](\xmin+0.2,\ymin+6.8) -- (\xmin+0.9,\ymin 5.2 );
  \draw[-,draw=black,very thick](\xmin,\ymin+9.8) -- (\xmin,\ymin 9.2 );
    \draw[-,draw=black,very thick](\xmin+0.2,\ymin+8.8) -- (\xmin+0.8,\ymin 8.2 );
    \draw[-,draw=black,very thick](\xmin-0.2,\ymin+8.8) -- (\xmin-0.8,\ymin 8.2 );
 \draw[-,draw=black,very thick](\xmin+0.8,\ymin+7.8) -- (\xmin+0.2,\ymin 7.2 );
    \draw[-,draw=black,very thick](\xmin-0.8,\ymin+7.8) -- (\xmin-0.2,\ymin 7.2 );
      \draw[-,draw=black,very thick](\xmin,\ymin+0.2) -- (\xmin,\ymin 0.8 );

      \draw[-,draw=black](\xmin,\ymin+6.8) -- (\xmin+0.1,\ymin 5.9 );
      \draw[-,draw=black](\xmin,\ymin+5.5) -- (\xmin-0.3,\ymin 5.2 );
      \draw[-,draw=black](\xmin+0.3,\ymin+5.5) -- (\xmin+0.7,\ymin 5.2 );

      \draw[-,draw=black](\xmin-0.1,\ymin+2.7) -- (\xmin-0.1,\ymin 2.55 );
            \draw[-,draw=black](\xmin-0.8,\ymin+2.05) -- (\xmin-0.3,\ymin 2.25 );
            \draw[-,draw=black](\xmin+0.7,\ymin+2.05) -- (\xmin+0.2,\ymin 2.25 );

\end{tikzpicture}
\end{center}
\caption{\small Closure diagram of the nilpotent orbits  of $SO(10,\cx)$ exhibiting the special orbits. The labels on the nodes represent partitions of $10$ and have to be completed by $1$s as needed, such that for example the label $3$ corresponds to the partition $3\cdot 1^7$. The two orbits in small font, $3\cdot 2^2\cdot1^3$ and $5\cdot 2^2\cdot1$, are not special. Removing them together with lines connected to them one obtains an `up-down' symmetric diagram as required by the Spaltenstein map.}
\label{ClosureDiagComplex}
\end{figure}

From the tensorial differential equation one derives that 
\bea 
\label{FCadj}
E_{\mbox{\DSOV{0}{s}000}}  &:&   |p|^2 |q|^2 - (p,q)^2 \ne 0 \quad \hspace{4mm} (p,q) \in \mathds{Z}^{12} \CR
 E_{\mbox{\DSOV{0}000s}} &:& |p|^2 = |q|^2 = (p,q)=0 \quad (p,q)  \in \mathds{Z}^{12}\cap\overline{\mathcal{O}}^9 \CR
 E_{\mbox{\DSOV{s}0000}}  &:& p\wedge q=0 \ \quad \hspace{21mm}(p,q)  \in \mathds{Z}^{12}\cap\overline{\mathcal{O}}^7 \CR
  E_{\mbox{\DSOV{\stfrac{3}{2}}0000}}  &:&|p|^2 = p\wedge q = |q|^2 = 0   \quad\hspace{1mm} (p,q)\in \mathds{Z}^{12} \cap\overline{\mathcal{O}}^6
 \eea
where  $\overline{\mathcal{O}}^n$ is the dimension $n$ algebraic variety in $\mathds{R}^{12}$ of solutions to the corresponding constraint. 
Here the value of $s$ is understood to be generic, whereas the support of the Fourier modes reduces at some specific values
\bea  E_{\mbox{\DSOV{0}0001}} &=& E_{\mbox{\DSOV{0}0010}} =\frac{1}{2} \frac{\zeta(3)}{\zeta(2)}  E_{\mbox{\DSOV{\stfrac{3}{2}}0000}} \CR
 E_{\mbox{\DSOV{0}{\stfrac{1}{2}}000}}&\propto&   E_{\mbox{\DSOV{1}0000}}  \CR
  E_{\mbox{\DSOV{0}{1}000}}&\propto&   E_{\mbox{\DSOV{0}0002}}=  E_{\mbox{\DSOV{0}0020}}  
\eea
where the differential equations degenerate.  Strictly speaking the undisplayed coefficients vanish as an artefact of Langlands normalisation, but we will see in the next subsection that these relations are indeed satisfied in a suitable normalisation in which the coefficients are finite. The value of $N$ is never constrained. When $N=0$, one can define the abelian Fourier modes $(p,q)$, and the reduction of a differential operator in a given representation implies the same constraint as the corresponding algebraic equation. We will moreover assume that the dimension of the algebraic variety that supports the non-abelian Fourier modes is still correctly determined by the algebraic solution to the constraint to be half the dimension of the corresponding nilpotent orbit, even if the structure of the solution is much more complicated in general for non-abelian Fourier modes.

Let us explain this in more detail in some examples. The vector Eisenstein series satisfies in general 
\be 
{\bf D}_{16}^{\; 2} E_{\mbox{\DSOV{\hspace{-0.0mm}{\mathnormal{s}}}0000}} = \frac{s(s-4)}{4} \mathds{1}_{16}  E_{\mbox{\DSOV{\hspace{-0.0mm}{\mathnormal{s}}}0000}} \ , \label{D162}  
\ee
and using the 3-graded  decompositions associated to \eqref{AdjointD5}
\bea 
{\bf 16} &\cong& {\bf 4}^\ord{-1} \oplus ( {\bf 2}, \overline{\bf 4})^\ord{0} \oplus {\bf 4}^\ord{1} \ , \CR
{\bf 10} &\cong& {\bf 2}^\ord{-1} \oplus {\bf 6}^\ord{0} \oplus {\bf 2}^\ord{1} \ , \label{3Grad} 
\eea
one computes that the restriction of \eqref{D162} to the degree $2$ component gives
\be 
\frac{1}{6} \varepsilon_{ipkl} \varepsilon^{\alpha\beta} Q_\alpha^{jp} Q_\beta^{kl} F_{Q,0\, \mbox{\DSOV{\hspace{-0.0mm}{\mathnormal{s}}}0000}}  = 0  \ \Rightarrow \ q\wedge p = 0 \ . 
\ee
Similarly for the spinor Eisenstein series 
 \be 
 {\bf D}_{10}^{\; 2}  E_{\mbox{\DSOV{0}000{\mathnormal{s}}}} = \frac{s(s-4)}{4} \mathds{1}_{10}  E_{\mbox{\DSOV{0}000{\mathnormal{s}}}}  \ , \label{SpinDiff} 
 \ee
one obtains the degree $2$ component 
\be  
\frac{1}{3} \varepsilon_{ijkl}  Q_\alpha^{ij} Q_{\beta}^{kl}    F_{Q,0\, \mbox{\DSOV{0}000{\mathnormal{s}}}}= 0 \  \Rightarrow \  \left( \begin{array}{cc}\ |q|^2\ &  (p,q) \\  (p,q) & \ |p|^2\   \end{array}\right) = 0 \ . \label{AlgDi} 
\ee
More generally Eisenstein series satisfy tensorial differential equations that are associated to the algebraic constraints defining the corresponding nilpotent orbits. Because the maximal degree component of these differential equations in a given parabolic gauge only involves the axion fields, they become algebraic constraints in Fourier space that reproduce the corresponding constraint. One can in this way understand the wave-front set from the differential equations satisfied by the Eisenstein series \cite{Bossard:2015uga}.  Strictly speaking the reduction of  \eqref{SpinDiff}  to the algebraic constraint \eqref{AlgDi} on the Fourier modes $(p,q)$ is only valid when $N=0$, whereas in general one gets a corresponding differential equation for the axion with a right-hand-side linear in $N$ that depends explicitly on the axion moduli. Nonetheless, the constraint still makes sense for counting the lattice dimension, \ie the dimension $n$ of the algebraic subvariety $\overline{\mathcal{O}}^n$ in the corresponding vector space of charges $\mathds{R}^{12}$ such that the Fourier modes are valued in $\mathds{Z}^{12} \cap\overline{\mathcal{O}}^n$. We can see this explicitly in the perturbative string theory limit, in which one decomposes instead the series in RR Fourier modes $e^{2\pi i ( p,a)}$, with the $Spin(4,4)$ Weyl spinor axion $a$. We compute in Appendix \ref{Poisson} using the Poisson summation formula that
\begin{multline}  
E_{\mbox{\DSOV{0}000s}} =\gs^{-s}  E_{\DSOIVS{0}{0}{0}{s}} + \gs^{s-4}   \frac{\xi(2(s-2))}{\xi(2s)} E_{\DSOIVS{0}{0}{s\mbox{-}1}{0}}\\ + \frac{2}{\xi(2s)} \sum_{\substack{p\in \mathds{Z}^8\\ (p,p)=0}} \Scal{{\rm gcd}(p)^{s-1} \sum_{n|{\rm gcd}(p) } n^{2(2-s)}}   \frac{\gs^{-2}}{|Z(p)|} E_{\DSLIII{s\mbox{-}1}00}(v_p) K_{s{-}2}(2\pi  \gs^{-1} |Z(p)|)e^{2\pi i ( p,a)}  \label{EDXspin} 
\end{multline}
where $v_p$ is the $SL(4)$ subgroup of $SO(4,4)$ that stabilises the null vector $p$, and $|Z(p)|^2$ is the $SO(4,4)$ moduli dependent invariant mass associated to the charge $p$. In this equation one finds that the RR Fourier modes are constrained to be null vectors, and the three extra integer sums come from the $SL(4)$ Fourier modes of the Eisenstein series in the fundamental representation, with total `lattice dimension' $10$, as for the ten-dimensional lattice associated to the spinor Eisenstein series in the M-theory limit. In this case however, we have a non-abelian decomposition for which each null vector $p$ of $\mathds{Z}^{8}$ defines a particular Fourier decomposition of the corresponding $SL(4)\subset SO(4,4)$ moduli space. The graded decomposition of the vector representation indeed implies that for a generic charge
\be
{\bf Q}_{10} = \left( \begin{array}{ccc} \ 0 \ & p & 0 \\
0 & \ {\bf Q}_8  \ & p \\
0 & 0 & 0  \ \end{array} \right) \in \mathfrak{so}(4,4) \oplus {\bf 8}^\ord{2}\subset \mathfrak{so}(5,5) \ , 
\ee
one has the quadratic constraint 
\be  
 {\bf Q}_{10}^{\; 2} = \left( \begin{array}{ccc} \ 0 \ &  {\bf Q}_8 p & (p,p) \\
0 & \ {\bf Q}_8^{\; 2}   \ &  {\bf Q}_8 p \\
0 & 0 & 0  \ \end{array} \right) \  = 0  \ , 
\ee
such that the $E_{\DSLIII{s\mbox{-}1}00}(v_p)$ must satisfy the quadratic differential equation associated to the minimal nilpotent orbit
\be
{\bf D}_6^{\; 2}E_{\DSLIII{s\mbox{-}1}00} =  \frac{(s-1)(s-3)}{4} \mathds{1}_{6} E_{\DSLIII{s\mbox{-}1}00} \ .  
\ee
For $s=1$, the  $SL(4)$ Eisenstein series is a constant, and one gets back $E{\mbox{\DSOV{\stfrac{3}{2}}{0}000}} $. 

Let us now consider the inhomogeneous equations (\ref{AnomaLaplace},\ref{EquaF1}). The Fourier modes of $\mathcal{E}_{\gra00}$  of vanishing D3-brane charge are defined by doublets of proportional null vectors and so its square involves the generic sum of doublets   $(p_i,q_i)$. One computes that the sum of such doublets gives generically a doublet of linearly independent non-null vectors  $(p,q)$ of opposite signature, that has a negative quartic $SL(2)\times SO(3,3)$ invariant 
\be   
|p|^2 |q|^2 - (p,q)^2 <0 \ \ . 
\ee
This can  easily be seen if we take 
\be 
(p_1,q_1) = (p,0) \ , \qquad (p_2,q_2) = (0,q) \ , \qquad (p,q) \ne 0 \ .
 \ee 
This corresponds precisely to the structure of the Fourier modes of a generic function $E{\mbox{\DSOV{0}{s}000}} $, although its Fourier modes can have the two signs for  $ |p|^2 |q|^2 - (p,q)^2$ in general.  To see this one uses the property that the `fundamental' representations are all 3-graded in this decomposition \eqref{3Grad} such that the positive degree generator in \eqref{AdjointD5} to the third power vanishes automatically in all these representations without any further restriction. Note that the real $SO(5,5)$ nilpotent orbit associated to instanton charges of negative quartic invariant corresponds to a complex $Sp(4,\mathds{C})\times Sp(4,\mathds{C}) $ orbit of a non-BPS solution through the Kostant--Sekiguchi correspondence \cite{CollingwoodMcGovern}. Nonetheless, a given representative does not need to lie in the intersection of the respective orbits that are related by the Kostant--Sekiguchi isomorphism. This additional restriction only applies when one requires moreover the corresponding black hole solution to be regular in seven dimensions \cite{Bossard:2009we}, but should not apply to the space-time instanton as such. 

One can also check the representation of the Fourier modes in the parabolic subgroup associated to the decompactification to eight dimensions 
\be 
\mathfrak{so}(5,5) \cong {\bf 3}^\ord{-4} \oplus ( {\bf 2}\otimes {\bf 2}\otimes \overline{\bf 3})^\ord{-2} \oplus \scal{ \mathfrak{gl}_1 \oplus \mathfrak{sl}_2\oplus \mathfrak{sl}_2\oplus \mathfrak{sl}_3}^\ord{0}  \oplus ( {\bf 2}\otimes {\bf 2}\otimes {\bf 3})^\ord{2} \oplus \overline{\bf 3}^\ord{4} \label{8DD5} \ , 
\ee
with 
\bea 
{\bf 16} &\cong& {\bf 2}^\ord{-3} \oplus ( {\bf 2}\otimes {\bf 3})^\ord{-1} \oplus ( {\bf 2}\otimes \overline{\bf 3})^\ord{1} \oplus {\bf 2}^\ord{3} \ , \CR
{\bf 10} &\cong& \overline{\bf 3}^\ord{-2} \oplus ({\bf 2}\otimes {\bf 2})^\ord{0} \oplus {\bf 3}^\ord{2} \ . 
\eea
The Fourier modes give a triplet of $SO(2,2)$ vectors $p_i$ associated to the effective Euclidean 1-brane coupled to the $6$ vector fields along $T^2$, and a conjugate triplet of singlets $q^i$ associated to the effective Euclidean 2-brane coupled to the $3$ tensor fields.\footnote{By effective Euclidean $p$-brane we mean any Euclidean $p+k$-brane that wrap $p$ directions of the decompactified $T^2$ and $k$ directions in the other $T^2$.}  For the function $E{\mbox{\DSOV{0}{s}000}} $, the condition that the differential operator in the spinor representation reduces to the third order implies 
\be 
\varepsilon^{ijk}  p_i \wedge  p_j \wedge p_k = 0 \ ,  
\ee
such that only two of the vectors are linearly independent. The counting of the number of modes here works as follows. There are two vectors ($4+4$) plus a third depending on the two others ($+2$) and the three weight $2$ components ($+3$), and we get back a $13$ dimensional lattice of Fourier modes. 

For the function $ E{\mbox{\DSOV{0}000s}}$, we moreover require the vector representation differential operator to reduce at second order, such that all the scalar products vanish
\be (p_i,p_j) = 0 \ ,  \ee
getting therefore three more constraints (since $p_3$ is already assumed to be a linear combination of $p_1$ and $p_2$), recovering the $10$ dimensional lattice of Fourier modes.

The minimal representation corresponds to the restriction in which all $p_i$ are proportional and null, and moreover $p_i q^i = 0$, such that one gets a $3+1+1+2=7$ dimensional lattice of Fourier modes as required. The generic sum of two such vectors gives a doublet of linearly independent vectors, but cannot give three linearly independent vectors, such that one gets back that the source term (\ref{AnomaLaplace},\ref{EquaF1}) indeed sources generic Fourier modes of a function of type $E{\mbox{\DSOV{0}{s}000}} $, and is consistent with it in the sense that it does not source more generic Fourier modes. 

\subsection{Whittaker vectors of $SO(5,5)$ series}

In this section we analyse the degenerate Whittaker vectors of some maximal parabolic $SO(5,5)$ Eisenstein series. Whittaker vectors are special cases of Fourier coefficients when the unipotent is taken to be maximal unipotent $N$ of all unipotent elements in a Borel subgroup. There is a close connection between Whittaker vectors and Fourier modes associated to nilpotent orbits~\cite{Ginzburg,Gustafsson:2014iva}.

The number of instanton charges that label Whittaker vectors is equal to the real rank of the group. For $SO(5,5)$ we therefore have five instanton charges $\vec{m}_N$ that can be arranged on the Dynkin diagram of $SO(5,5)$. When some of the charges vanish, Whittaker vectors are called degenerate and a general formalism for determining degenerate Whittaker vectors was presented in~\cite{Fleig:2013psa}. We will write a Whittaker vector as
\begin{align}
W_{\DSOV{s_1}{s_2}{s_3}{s_4}{s_5}}\left(\DSOV{m_1}{m_2}{m_3}{m_4}{m_5}\right) =
\int\limits_{N(\ints)\backslash N(\reals)} E_{\DSOV{s_1}{s_2}{s_3}{s_4}{s_5}} \overline{\psi_{\DSOV{m_1}{m_2}{m_3}{m_4}{m_5}}} \ , 
\end{align}
with $\psi{\DSOV{m_1}{m_2}{m_3}{m_4}{m_5}}$ defined as in \eqref{PsiFourier}. The Whittaker function types can be labelled by the same labels as the (complex) nilpotent orbits and we will refer to them by these names. The wave-front set of an Eisenstein series will be the largest complex orbit with non-vanishing Whittaker vectors, which is unique \cite{Joseph,BarbaschVogan}.\footnote{\ie the closure of this orbit includes all the orbits for which the Whittaker vectors do not vanish.} In the following we will always evaluate the Whittaker vectors at the origin of moduli space, \ie the identity of the Cartan subgroup.

\subsubsection{$SO(5,5)$ vector series}
\label{sec:D5vec}

The series
\begin{align}
\label{D5vec}
E_{\DSOV{s}{0}{0}{0}{0}} =\frac{\xi (2 s-6) \xi (2 s-4)}{\xi (2 s) \xi (2 s-3)} E_{\DSOV{4-s}{0}{0}{0}{0}} 
\end{align}
can have Fourier coefficients at most associated with the orbit of type $(2A_1)'$ since it can be functionally realised on the coset $P_1\backslash SO(5,5)$ of dimension $8$. The functional dimension of an automorphic realisation is half the dimension of the maximal nilpotent orbit contributing to the Fourier coefficients and therefore inspection of table~\ref{tab:D5orbs} shows that there can be no Fourier coefficients beyond the orbit $(2A_1)'$. 

The Whittaker vectors corresponding to the $(2A_1)'$ orbit can be represented by instanton charges with Dynkin diagram $
\DSOV{0}{0}{0}{m}{n} $
for $m,n\in\ints$. By contrast, the instanton charges for the $(2A_1)''$ orbit are for example of the type $\DSOV{m}{0}{n}{0}{0}$. One can check that the associated Whittaker vectors vanish for all values of $s$ in~\eqref{D5vec}. For a generic $s$ in~\eqref{D5vec}, the Whittaker vector can be evaluated with the help of~\cite{Fleig:2013psa} to be
\begin{align}
\label{vecD5:2A1}
W_{\DSOV{s}{0}{0}{0}{0}}\left({\DSOV{0}{0}{0}{m}{n}}\right) = \frac{\xi(2s-3)}{\xi(2s)} W_{\DSLI{s-\frac32}}\left({\scriptstyle [m]}\right)W_{\DSLI{s-\frac32}}\left({ \scriptstyle [n]} \right),
\end{align}
where we have placed ourselves at the origin in moduli space for simplicity. Here, the function
\begin{align}
W_{\DSLI{s}}({\scriptstyle [n]}) = \frac{2}{\xi(2s)} |n|^{s-\frac{1}{2}}  \Scal{ \sum_{k|n} k^{1-2s} }K_{s-1/2} (2\pi |n|) 
\end{align}
is the Whittaker function for the $SL(2)$ series $E_{\scriptstyle [{s}]}$ at the origin of moduli space. It vanishes linearly for $s\to 0$ (and $s\to\tfrac12$) due to the $\xi(2s)$ denominator. Inspecting~\eqref{vecD5:2A1} then shows that the whole Whittaker vector for the $SO(5,5)$ vector series vanishes for $s\to \frac32$. 

Performing a similar calculation for the smaller $A_1$ orbit shows that the associated Whittaker vector never vanishes (as a function of $s\ne 0$) confirming the reduction of the Fourier coefficients already mentioned above. 

\subsubsection{$SO(5,5)$ adjoint series}
\label{sec:D5adj}

The adjoint Eisenstein series of $SO(5,5)$
\begin{align}
\label{D5adj}
E_{\DSOV{0}{s}{0}{0}{0}} = \frac{\xi (2 s-6) \xi (2 s-5) \xi (2 s-4) \xi (4 s-7)}{\xi (2 s) \xi (2 s-2) \xi (2 s-1) \xi (4 s-6)} E_{\DSOV{0}{\frac72-s}{0}{0}{0}}
\end{align}
can be realised as a function on the space $P_2\backslash SO(5,5)$ of dimension $13$. The maximal nilpotent orbit supported by this function for generic $s$ is therefore the orbit of dimension $26$, labelled type $A_2$ in table~\ref{tab:D5orbs}.

For special values of $s$ one has reductions of the orbit type. As indicated in~\eqref{D5adj}, the adjoint series has a functional relation that relates $s\leftrightarrow \frac72-s$. In Langlands normalisation, one can use the Langlands constant term formula to show that $E\DSOV{0}{s}{0}{0}{0}$ has simple zeroes at $s=\frac12$, $s=1$ and $s=\tfrac32$. Inspecting the prefactor in the functional relation one concludes that there has to be a simple pole for $s=2$, $s=\frac52$ and $s=3$. These have to be taken into account when discussing the simplifications in the Whittaker vectors.

Performing this analysis here implies that the degenerate Whittaker vector of type $A_2$ takes the value
\begin{align}
\label{adjD5:A2}
W_{\DSOV{0}{s}{0}{0}{0}}\left({\DSOV{1}{1}{0}{0}{0}}\right) = \frac{\xi(2s-3)\xi(2s-4)}{\xi(2s)\xi(2s-2)} W_{\DSLII{s-1/2}{s-2}}\left({\scriptstyle [1\, 1]}\right),
\end{align}
for unit charges. Let us discuss various limits of this formula.
\begin{itemize}
\item In the limit $s\to \frac12$, the $SL(3)=A_2$ Eisenstein series becomes of minimal type~\cite{Pioline:2009qt} and the corresponding Whittaker vector vanishes linearly as does the prefactor. This means that the suitably normalised adjoint $SO(5,5)$ Eisenstein series has no $A_2$ Whittaker vectors for $s=\frac12$. This is consistent with the fact that it is related to a different Eisenstein series through
\begin{align}
E_{\DSOV{0}{s}{0}{0}{0}} = \frac{\xi(2s-2)}{\xi(2s)} E_{\DSOV{\frac32-s}{0}{s-\frac12}{0}{0}}.
\end{align}
This shows that in the limit $s\to\frac12$ the suitably normalised adjoint series is proportional to the vector series at $s=1$ and therefore has Whittaker vectors of type $(2A_1)'$ (and smaller).

\item In the limit $s\to 1$, the prefactor in~\eqref{adjD5:A2} vanishes linearly and so does the Whittaker vector on the right-hand side, giving a total vanishing up to quadratic order. In view of the linear vanishing of the Eisenstein series in this limit this means that the adjoint $SO(5,5)$ series has no $A_2$ Whittaker vectors in the limit $s\to 1$. Turning to the $3A_1$ Whittaker vectors shows that they also vanish in this limit. Further analysis shows that there is an effective reduction to type  $(2A_1)''$ for the Fourier coefficients. The adjoint series indeed satisfies the functional relation
\begin{align}
E_{\DSOV{0}{s}{0}{0}{0}} = \frac{\xi (2 s-4) \xi (2 s-3) \xi (4 s-7)}{\xi (2 s) \xi (2 s-1) \xi (4 s-6)} E_{\DSOV{0}{0}{s-1}{4-2s}{0}},
\end{align}
showing that at $s=1$ there is a relation to the chiral spinor series that will be discussed below.

\item In the limit $s\to \frac32$,  the $SL(3)=A_2$ Eisenstein series becomes of minimal type~\cite{Pioline:2009qt} and the corresponding Whittaker vector vanishes linearly as above. But in this case the prefactor remains finite. As the whole Eisenstein series vanishes for $s\to\frac32$, there is in fact no simplification in the degenerate Whittaker vector if one removes the overall vanishing by suitable normalising $\xi$ factor. If one performs a similar analysis for the $3A_1$ type Whittaker vectors one also concludes that there is no simplification in this limit. Therefore, the suitably normalised $s=\frac32$ adjoint Eisenstein series has Whittaker vectors of all types up to $A_2$ and indeed is of the same type as the generic adjoint Eisenstein series.  

However, the $s=\tfrac32$ adjoint function differs from the generic case in that the Eisenstein series $E{\DSOV{0}{\tfrac32}{0}{0}{0}}$ is square integrable according to Langlands' criterion~\cite[\S5]{Langlands}. This is similar to what happens for the minimal series (for $D\leq 6$) and the  next-to-minimal series (for $D\leq 4$) \cite{Green:2011vz} and signals that it belongs to a small unitary representation. The generating character of the adjoint Eisenstein series at $s=\frac32$ does satisfy to an additional quartic differential constraint associated to the $3A_1$ nilpotent orbit, and by Langlands formula all its non-vanishing constant terms (in $e^{\langle w\lambda + \rho,H(\cV)\rangle}$) do as well, although its Fourier modes violate this constraint as we have just exhibited. Moreover, the Laplace eigenvalue is $-12$ and thus outside the range of the associated continuous (degenerate) principal series of solutions to \eqref{eq:DEadjD5} with $s=\frac74+i r$ ($r\in\reals$) that has Laplace eigenvalues given by $-\frac{49}{4}-4r^2$. It therefore is part of the discrete spectrum of the Laplace operator.\footnote{An analogous phenomenon happens for the adjoint $E_{7(7)}$ series at $s=4$ that was mentioned at the end of section~\ref{sec:DE}. }
\end{itemize}

Note that the Whittaker vector only spans the Fourier modes associated to the non-BPS real nilpotent orbit of $SO(5,5)$. The notation $A_2$ means that a representative of the nilpotent orbit can be defined as a linear combination of the simple roots of a subalgebra $\mathfrak{sl}_3\subset \mathfrak{so}(5,5)$. In this case one can indeed realise a representative of the nilpotent orbit of dimension 26  as a linear combination of the two simple roots of $\mathfrak{sl}
_3$ through the embedding 
\be SL(3) \subset  SO(1,1) \times SL(3) \times SO(2,2) \subset SO(3,3) \times SO(2,2)  \subset SO(5,5)\ ,\ee so that the Levi subgroup of the real stabilizer of the corresponding nilpotent orbit is $SO(1,1)\times SO(2,2)$. One straightforwardly works out that this is also the stabilizer of a doublet of linearly independent non-null vectors of opposite signature in $SL(2) \times SO(3,3)$. However, a doublet of linearly independent non-null vectors of the same signature is stabilized by $SO(2) \times SO(3,1)$, which is the centralizer subgroup of $SU(2,1)$ in $SO(5,5)$, through the embedding 
\be 
SU(2,1)\subset SO(2)\times SU(2,1) \times SO(1,3)\subset SO(4,2) \times SO(1,3) \subset SO(5,5) \ .
\ee 
Nonetheless, it seems clear that the adjoint Eisenstein series should have non-zero modes in all these three real orbits, as expected from the discussion in~\cite{Vogan}.

\subsubsection{$SO(5,5)$ spinor series}

The chiral spinor series
\begin{align}
\label{spinD5}
E_{\DSOV{0}{0}{0}{s}{0}} = \frac{\xi(2s-5)\xi(2s-7)}{\xi(2s)\xi(2s-2)} E_{\DSOV{0}{0}{0}{0}{4-s}}
\end{align}
is related to the anti-chiral spinor series as shown. It can be realised through functions on the space $P_4\backslash SO(5,5)$ of dimension $10$. Referring back to table~\ref{tab:D5orbs} we conclude that at most the type $(2A_1)''$ orbit can appear in the Fourier coefficients. The series $E\DSOV{0}{0}{0}{s}{0}$ has simple zeroes for $s=\frac12$ and $s=\frac32$.

The type $(2A_1)''$ Whittaker vector is
\begin{align}
\label{spinD5:2A1pp}
W_{\DSOV{0}{0}{0}{s}{0}}\left({\DSOV{n}{0}{m}{0}{0}}\right) = \frac{\xi(2s-3)\xi(2s-3)}{\xi(2s)\xi(2s-2)} W_{\DSLI{s-\frac32}}\left({\scriptstyle [n]}\right)W_{\DSLI{s-\frac32}}\left({\scriptstyle [m]}\right)
\end{align}
Note that this formula is perfectly consistent with \eqref{EDXspin}, identifying the Whittaker vector of the $SL(4)$ function 
\be W_{\DSLIII{s-1}{0}{0}}\left({\scriptstyle [0,n,0]} \right) =   \frac{\xi(2s-3)}{\xi(2s-2)} W_{\DSLI{s-\frac32}}\left({\scriptstyle [n]}\right) \ , \ee
one is left with a remaining factor of  $\frac{\xi(2s-3)}{\xi(2s)} W_{\DSLI{s-\frac32}}\left({\scriptstyle [{\rm gcd}(p)]}\right) $ for the integral null vector $p$ that matches precisely \eqref{EDXspin}.

We again discuss various limiting values of~\eqref{spinD5:2A1pp} for $s$.
\begin{itemize}
\item In the limit $s\to \frac12$, the prefactor in~\eqref{spinD5:2A1pp} vanishes linearly but the two $SL(2)$ Whittaker functions remain finite. This is in agreement with the linear vanishing of the whole Eisenstein series and means that a suitably normalised version does not exhibit any simplifications in this limit.

\item In the limit $s\to 1$, the prefactor vanishes linearly and the two Whittaker functions in~\eqref{spinD5:2A1pp} remain finite, implying an overall linear vanishing of the $(2A_1)''$ Whittaker function~\eqref{spinD5:2A1pp} for the $SO(5,5)$ chiral spinor series (since the Eisenstein series itself is regular at this value). Indeed, there is a functional relation
\begin{align}
E_{\DSOV{0}{0}{0}{s}{0}} = \frac{\xi(2s-4)}{\xi(2s)} E_{\DSOV{\frac52-s}{0}{0}{0}{s-1}},
\end{align}
showing that in the limit $s\to 1$, the chiral spinor series will have the same behaviour as the vector series~\eqref{D5vec} at $s=\frac32$. We already showed in section~\ref{sec:D5vec} above that in this case one has a reduction to the minimal $A_1$ orbit.

\item In the limit $s\to \frac32$, the prefactor in~\eqref{spinD5:2A1pp} diverges linearly whereas the two Whittaker vectors each vanish linearly, giving a linearly vanishing result in agreement with the behaviour of the chiral spinor series~\eqref{spinD5} at $s=\frac32$. This implies that there is no simplification in the Fourier coefficients for $s\to\frac32$.

\item In the limit $s\to 2$ the Whittaker vector~\eqref{spinD5:2A1pp} remains finite and therefore there is no simplification in the Fourier coefficients and the series is of type $(2A_1)''$. This is the case that is related to adjoint function at $s=1$.

\end{itemize}

Let us also note that the $(2A_1)'$ type Whittaker vectors {\em always} vanish, consistently with the analysis of the preceding section. In summary, the chiral spinor series~\eqref{spinD5} always has Fourier coefficients attached to the orbits $(2A_1)''$ and $A_1$ (and, of course, the trivial orbit) except for $s=1$  when the Fourier coefficients reduce to just type $A_1$. 

\subsection{Relation to supersymmetric invariants}

Returning to the $\nabla^6 R^4$ threshold function discussed in this paper and given in~\eqref{E01D6}, we conclude therefore that the function $E{\DSOV{0}{0}{0}{0}{4}}$ appearing in $\cE_{\gra01}^{\ord{6}}$ only gets corrections  associated to $\tfrac14$-BPS corrections. Although the adjoint Eisenstein series $E{\DSOV{0}{\frac72}{0}{0}{0}}$ does include constant terms inconsistent with perturbative string theory, its  wave-front set is of type $A_2$, consistent with the property that the $\nabla^6 R^4$ threshold function gets corrections associated to $\tfrac18$-BPS instantons.  

The $\nabla^4 R^4$ threshold function $\cE_{\gra10}^{\ord{6}}$ is given by a combination of a spinor and vector function by~\cite{Green:2010kv,Pioline:2015yea}
\begin{align}
\cE^{\ord{6}}_{\gra01} = \zeta(5) \hat{E}_{\DSOV{\tfrac52}{0}{0}{0}{0}} + \frac{8\zeta(6)}{45} \hat{E}_{\DSOV{0}{0}{0}{0}{1}}.
\end{align}
From the analysis of the Whittaker vectors and Fourier modes we now see that the two functions have wave-front sets of types $(2A_1)'$ and $(2A_1)''$, respectively, that correspond to the two distinct supersymmetric $\tfrac14$-BPS invariants~\cite{Bossard:2014aea}. The general pattern seems to be that $\tfrac14$-BPS corrections should be associated with all (special) $2A_1$-type orbits. 

\subsection*{Acknowledgements}
We thank Daniel Persson and Boris Pioline for useful discussions. G.B. is grateful to AEI for its kind hospitality where this work was initiated. A.K. is grateful to IHES for its kind hospitality where part of this work was carried out. 

\appendix

\section{Adjoint $E_{8(8)}$ series as a lattice sum}
\label{E8Diff} 
In this appendix, we rewrite the adjoint Eisenstein series~\eqref{E8adj} as a sum over a $248$-dimensional lattice in the adjoint representation. For this we recall that the symmetric tensor product of the ${\bf 248}$ of $E_{8(8)}$ is
\begin{align}
{\bf 248} \otimes_s {\bf 248} = {\bf 27\,000} \oplus {\bf 3875} \oplus {\bf 1}.
\end{align}
The minimal representation-theoretic constraint to be imposed on a lattice sum is that a charge $Q\in\ints^{248}\subset \reals^{248}$ satisfy the constraint that its square contain only the largest representation ${\bf 27\,000}$, as is necessary for a lattice sum to be automatically an eigenfunction of the Laplacian~\cite{Obers:1999um}.

Decomposing the adjoint ${\bf 248}$ into ${\bf 128}\oplus{\bf 120}$ under the $Spin(16)$ subgroup, one can write moduli-dependent $\mathfrak{e}_8$-valued charge $\cV Q \cV^{-1}$ (with $\cV$ the $248$-bein) in terms of a Majorana--Weyl spinor $X_A$ (with $A=1,\ldots,128$) and the antisymmetric tensor $\Lambda_{ij}=\Lambda_{[ij]}$ (with $i,j=1,\ldots,16$) of $\mathfrak{so}(16)$. The invariant mass is then given by $XX=X_A X^A$. The representation-theoretic constraint that $Q\otimes Q$ must have no component in the  ${\bf 1}\oplus {\bf 3875}$ reads
\be  
\Lambda_{ik} \Lambda^{jk} = \frac{1}{16} \delta_i^j (XX) \ , \qquad \Lambda_{ij} \Gamma^j X = \frac{1}{16} \Gamma_i\,  \baa \Lambda X \ , \qquad \Lambda_{[ij} \Lambda_{kl]} = - \frac{1}{48} ( X \Gamma_{ijkl} X) \ . 
\ee
By construction the derivative $\cD_A$ acts on these tensors in the adjoint representation of $\mathfrak{e}_{8(8)}$, \ie
\be 
\cD_A X_B = \frac{1}{4} \baa \Lambda_{AB} \ , \qquad \cD_A \Lambda_{ij} = \frac{1}{4} \Gamma_{ij\, AB} X^B \ . 
\ee
Using among other equations that 
\be 
 \baa \Lambda \Gamma_{ijkl} \, \baa \Lambda\, =  - 48 \Lambda_{[ij} \Lambda^{pq} \Gamma_{kl]pq}   - \frac{3}{2} (X \Gamma_{[ij}{}^{pq} X) \Gamma_{kl]pq} - \frac{1}{2} (X \Gamma_{ijkl} X) - \frac{1}{48}  (X \Gamma^{pqrs} X) \Gamma_{ijklpqrs} \ , 
 \ee
one computes that 
\be 
\Delta (XX)^{-s} = \cD_A \cD_A (XX)^{-s} = 2s(2s-29) (XX)^{-s} \ , 
\ee
and
\bea \label{DerivativeEs} ( \cD \Gamma_{ijkl} \cD) (XX)^{-s} &=& 2s(2s-5) ( X \Gamma_{ijkl} X) (XX)^{-s-1} \ , \\
\Gamma^{kl} \cD ( \cD \Gamma_{ijkl} \cD) (XX)^{-s} &=& 2s(2s-5) \scal{ -48(2s-9) \Lambda_{ij} + (s-15) \Gamma_{ij} \baa \Lambda} X (XX)^{-s-1} \ , \CR
(\cD \Gamma_{ij} \Gamma_{pq} \cD) ( \cD \Gamma^{klpq} \cD) (XX)^{-s} &=& 8s(2s-5) \bigl(  96(2s-9)(s-7)  \Lambda_{ij} \Lambda^{kl}\bigr . \CR&& \bigl . \hspace{-30mm}  - \scal{ s(2s-41) + 102} ( X\Gamma_{ij}{}^{kl} X)  + 2(2s-21)(s-22) \delta_{ij}^{kl} (XX) \bigr)  (XX)^{-s-1}\ ,  \CR
(\cD \Gamma_{i[jk}{}^r \cD) ( \cD \Gamma_{lpq]r} \cD) (XX)^{-s} &=& -2s(2s-5) \scal{ 2s(2s-29)+48} \delta_{i[j}  \scal{ X \Gamma_{klpq]} X} (XX)^{-s-1}\,  . \nn \eea
We give some indications of how one derives these relations. First, one straightforwardly checks that 
\be \Gamma^{kl} \cD ( \cD \Gamma_{ijkl} \cD) (XX)^{-s} = s \scal{ 240  a(s) \Lambda_{ij} +  b(s) \Gamma_{ij} \baa \Lambda } X (XX)^{-s-1}\ , \ee
using representation theory and the ${\bf 3875}$ constraint, but computing the explicit coefficients $a(s)$ and $b(s)$ is rather cumbersome. To do so we use the property that the projector to the $\WSOXVI01000001$ is defined as
\be \scal{ \Gamma^{kl} \cD ( \cD \Gamma_{ijkl} \cD)  +  \Gamma_{ij} \cD \scal{ \tfrac{14}{5} \Delta + 336}} (XX)^{-s} = s a(s) \scal{ 240 \Lambda_{ij} +  \Gamma_{ij} \baa \Lambda } X (XX)^{-s-1}\ . \ee
This determines $b(s)$. Using moreover the general identity 
\be\Gamma_{[ij}  \scal{ \Gamma^{pq} \cD ( \cD \Gamma_{kl]pq} \cD)  +  \Gamma_{kl]} \cD \scal{ \tfrac{14}{5} \Delta + 336}}  = - 16 ( \cD \Gamma_{ijkl} \cD) \scal{  \tfrac{1}{5} \Delta + 180 } \ , \ee
one determines $a(s)$. Once the second identity in \eqref{DerivativeEs} is known, it is straightforward to compute the third one. The last one is more complicated to obtain, but one can nonetheless straightforwardly check that $(\cD \Gamma_{i[jk}{}^r \cD) ( \cD \Gamma_{lpq]r} \cD) (XX)^{-s} $ reduces to the product of a tensor quartic in $X$ times $(XX)^{-s-2}$. There is a unique quartic tensor in the $\WSOXVI10001000$, and a unique quartic tensor in the  $\WSOXVI00010000$, and using the ${\bf3875}$ constraint, one obtains that the former vanishes 
\be (X \Gamma_{i[jk}{}^r X) ( X \Gamma_{lpq]r} X) = - \delta_{i[j} ( X \Gamma_{klpq]} X) (XX) \ , \ee
such that 
\be (\cD \Gamma_{i[jk}{}^r \cD) ( \cD \Gamma_{lpq]r} \cD) (XX)^{-s}   =  c(s)  \delta_{i[j} ( X \Gamma_{klpq]} X) (XX)^{-s-1} \ , \ee
for a coefficient $c(s)$. The latter is fixed using the projector to the $\WSOXVI10001000$ of the fourth order differential operator,
\be \cD^4_{\mbox{\WSOXVI10001000}}  =  \scal{ \cD \Gamma_{i[jk}{}^r \cD} \scal{ \cD \Gamma_{lpq]r} \cD }+ \delta_{i[j} \scal{ \cD \Gamma_{klpq]} \cD} ( \Delta+48  )  \ . \ee

One finds in this way that the 1/2 BPS equation required for a $R^4$ type invariant is satisfied for $s=\frac{5}{2}$, 
\be (\cD \Gamma_{ijkl} \cD) (XX)^{-\frac{5}{2}} = 0 \ , \ee
while the 1/4 BPS equation required for a $\nabla^4 R^4$ type invariant is satisfied for $s=\frac{9}{2}$, 
\be \Gamma^{kl} \cD ( \cD \Gamma_{ijkl} \cD) (XX)^{-\frac{9}{2}} = -168 \Gamma_{ij} \cD  (XX)^{-\frac{9}{2}} \ , \ee
and the 1/8 BPS equation required for a $\nabla^6 R^4$  type invariant is satisfied for $s=\frac{11}{2}$, 
\be (\cD \Gamma_{i[jk}{}^r \cD) ( \cD \Gamma_{lpq]r} \cD) (XX)^{-\frac{11}{2}} = 150 \delta_{i[j}  (\cD \Gamma_{klpq]} \cD) (XX)^{-\frac{11}{2}}    \ .\ee 
For any $s$ the function $(XX)^{-s}$ satisfies that the fourth derivative restricted to the $\WSOXVI10001000$ irreducible representation vanishes. For $s=\frac{5}{2}$, the second derivative restricted to the $\WSOXVI00010000$ vanishes, for $s=\frac{9}{2}$ the third derivative restricted to the $\WSOXVI01000001$ vanishes, and for $s=7$ the fourth derivative restricted to the  $\WSOXVI02000000$ vanishes. 

By analytic continuation, the quartic constraint in the $\WSOXVI10001000$ is also satisfied by the Eisenstein series 
\be 
E_{\DEVIII{0}{0}{0}{0}{0}{0}{0}{s}} = \frac{1}{2\zeta(2s)}  \sum_{\vspace{-2mm}\begin{array}{c}\scriptstyle \vspace{-4mm} Q \in \mathds{Z}^{248} \vspace{2mm}\\ \scriptscriptstyle Q\times Q|_{\bf 3875}=0\end{array}}  (X(Q)X(Q))^{-s} \ ,
\ee
for almost all $s$. The normalisation here expresses the fact that there is one $E_8(\ints)$ $\tfrac12$-BPS orbit on $\ints^{248}$ for every $k\in\ints$.
 The series only converges for $s> 29$, but it is defined for complex $s$ by Langlands and all the values we consider here are regular. We know that the quadratic constraint and the cubic constraint are indeed satisfied by the Eisenstein series at $s=\frac{5}{2}$ and $s=\frac{9}{2}$, respectively \cite{Green:2011vz}. However, a more careful analysis exhibits that the Eisenstein series does not satisfies the quartic constraint in the $\WSOXVI02000000$  at $s=7$, and the wave-front set does not reduce at this specific value. One understands mathematically this property because the corresponding nilpotent orbit is not special in this last case. We also note that the adjoint $E_{8(8)}$ series at $s=7$ is square integrable and part of the discrete spectrum of the Laplace operator, similar to the adjoint $SO(5,5)$ series analysed in detail in section~\ref{sec:D5adj}.

\section{Poisson summation for the $SO(5,5)$ spinor series}
\label{Poisson}
In the string perturbation theory limit, the function $E{\mbox{\DSOV{0}000s}}$ defined as a sum over pure spinors in $\mathds{Z}^{16}$ (integers points of an eleven-dimensional variety), decomposes as a sum over chiral  and antichiral spinors $q$ and $p$ of $Spin(4,4)$ that have a vanishing norm and satisfy $(q\Gamma p)=0$. This gives the sum 
\be E_{\mbox{\DSOV{0}000s}} =e^{-s \phi}  E_{\DSOIVS{0}{0}{0}{s}} + \frac{1}{2\xi(2s)} \sum_{\substack{p\in \mathds{Z}^8\\ (p,p)=0}} \sum_{\substack{q\in \mathds{Z}^8\\ (p\Gamma q)=0}}  \int_0^\infty \frac{dt}{t^{1+s}} e^{ - \frac{\pi}{t}\scal{ e^{\phi} |Z(q+\ba \, a p)|^2 + e^{-\phi} |Z(p)|^2}} \ee
where the lattice sum is divided by $2\zeta(2s)$ to get the Langlands normalisation, and $e^\phi=g$ is the six-dimensional effective string coupling constant. One can always find an element of  $Spin(4,4,\mathds{Z})$ to rotate the null spinor $p$ to a preferred basis decomposing as
\be {\bf 8}_a = {\bf 1}^\ord{-2} \oplus {\bf 6}^\ord{0}\oplus {\bf 1}^\ord{2} \ , \quad  {\bf 8}_c = {\bf 4}^\ord{-1} \oplus \overline{\bf 4}^\ord{1} \ ,\quad  {\bf 8} = \overline{\bf 4}^\ord{-1} \oplus {\bf 4}^\ord{1} \ , \label{Pref1} \ee
with respect to $SL(4) \subset Spin(4,4)$, such that 
\be p = (0,0,n_p)\ , \quad q = (0,q_4) \ , \ee
where $n_p\in \mathds{Z}$, and $q_4\in \mathds{Z}^4$. One can therefore use the Poisson summation formula to exchange the sum over $q$ with a sum over the dual spinor $\tilde{q} = ( \tilde{q}_4,0)$. Note that the axion decomposes as required for the Poisson formula to disentangle the axion dependence as usual, because for $a = ( a_4,\tilde{a}_4)$ one has $\ba a p = ( 0 , n_p a_4)$ so that $q + \ba a p$ indeed defines a four vector of $SL(4)$, and $(p \ba a \tilde q) = {\rm gcd}(p) ( \tilde q_4,a_4)$. It remains to compute the determinant for the quadratic term in $q$. By construction, the $SL(4)$ factor vanishes and one gets the square of the normalised mass square of $p$,  $\scal{\frac{|Z(p)|}{n_p}}^2$, such that 
\bea E_{\mbox{\DSOV{0}000s}} &=&e^{-s \phi}  E_{\DSOIVS{0}{0}{0}{s}} + \frac{e^{-2\phi} }{2\xi(2s)}\hspace{-2mm} \sum_{\substack{p\in \mathds{Z}^8\\ (p,p)=0}}\sum_{\tilde{q}\in \mathds{Z}^4}  \int_0^\infty\hspace{-1mm} \frac{dt}{t^{s-1}} \frac{n_p^{\; 2}}{|Z(p)|^2} e^{ - \frac{\pi}{t}  e^{-\phi} |Z(p)|^2 - \pi t e^{-\phi} |Z(\tilde{q})|^2 + 2\pi i (p \ba\,  a \tilde q)} \CR
&=& e^{-s \phi}  E_{\DSOIVS{0}{0}{0}{s}} + e^{(s-4)\phi}  \frac{\xi(2(s-2))}{\xi(2s)} E_{\DSOIVS{0}{0}{s\mbox{-}1}{0}} \CR
&& \ + \frac{e^{-2\phi}}{\xi(2s)} \sum_{\substack{p\in \mathds{Z}^8\\ (p,p)=0}} \sum_{\tilde{q}\in \mathds{Z}^4_*}  n_p^{\; 2} \frac{|Z(\tilde{q})|^{s-2}\hspace{-3mm}}{|Z(p)|^s}\hspace{2mm} K_{s-2}(2\pi  e^{-\phi} |Z(p\Gamma\tilde{q})|)e^{2\pi i  (p \ba\,  a \tilde q)}  \ ,
\eea
where we have
 \be |Z(p\Gamma\tilde{q})| = |Z(p)| |Z(\tilde{q})| \ . \ee
At this point it is convenient to use triality, to define the equivalent decomposition
\be {\bf 8}_a =  \overline{\bf 4}^\ord{-1} \oplus {\bf 4}^\ord{1}   \ , \quad  {\bf 8}_c = \ {\bf 4}^\ord{-1} \oplus \overline{\bf 4}^\ord{1} \ , \quad  {\bf 8} =  {\bf 1}^\ord{-2} \oplus {\bf 6}^\ord{0}\oplus {\bf 1}^\ord{2} , \label{Pref2}\ee
such that
\be p = (0,p_4)\ , \quad \tilde{q} = (0, \tilde{q}^\prime_4) \ ,  \quad  (p\Gamma\tilde{q}) =(0,0,(p_4 \tilde{q}_4^\prime))\ ,  \ee
and \be  \frac{|Z(\tilde{q})|^{s-2}\hspace{-3mm}}{|Z(p)|^s}\hspace{2mm}  = {\rm gcd}(p\Gamma\tilde{q})^{s-1}  \frac{|v(p_4)|^{-2(s-1)}\hspace{-10mm}}{|Z(p\Gamma\tilde{q})|} \hspace{10mm} \ , \ee
where $|v(p_4)|$ is the invariant norm associated to the $SL(4)$ subgroup stabilising $(\tilde{q}\Gamma p)$. The sum can be replaced by the $P_3(\mathds{Z})\backslash SO(4,4,\mathds{Z})$ coset sum rotating $p$ to the preferred basis \eqref{Pref1}, the $P_1(\mathds{Z})\backslash SL(4,\mathds{Z})$ coset sum rotating $\tilde{q}_4$ to a preferred basis and the sum over their respective relative integer greatest common divisors. In this basis, $(p\Gamma\tilde{q}) = ( 0 , n_p \tilde{q}_4)$,  the $P_1(\mathds{Z})\backslash SL(4,\mathds{Z})$ coset element equivalently determines the direction of $(p\Gamma\tilde{q})$ as a vector. It follows by triality that one can rewrite this sum as the $P_1(\mathds{Z})\backslash SO(4,4,\mathds{Z})$ coset sum rotating $(p\Gamma\tilde{q})$ to the preferred basis \eqref{Pref2} and   the $P_1(\mathds{Z})\backslash SL(4,\mathds{Z})$ coset sum rotating $p_4$ to a preferred basis, together  with the sum over their respective relative integer greatest common divisors, keeping in mind that the greatest common divisor of $p$ divides $(p\Gamma\tilde{q})$. The $P_1(\mathds{Z})\backslash SO(4,4,\mathds{Z})$ coset sum together with the sum over the $(p\Gamma\tilde{q})$ greatest common divisor then reduces to the sum over all integral null vectors $(p\Gamma\tilde{q})$, while
\be \sum_{\gamma\in P_1(\mathds{Z})\backslash SL(4,\mathds{Z})}2 \sum_{n_p |{\rm gcd}(p\Gamma\tilde{q})}   n_p^{\; 2} |v(p_4)|^{-2(s-1)}=  \sum_{n_p |{\rm gcd}(p\Gamma\tilde{q})}  2 n_p^{\, 2(2-s)} E_{\DSLIII{s\mbox{-}1}00}(v_{(p\Gamma\tilde{q})})\ ,  \ee
where the factor of 2 appears because the sum over $n_p$ is then restricted to the positive integers dividing  $(p\Gamma\tilde{q})$. 
Finally, one obtains after renaming $(p\Gamma\tilde{q})$ as $p$ for simplicity, 
\begin{multline}  E_{\mbox{\DSOV{0}000s}} =e^{-s \phi}  E_{\DSOIVS{0}{0}{0}{s}} + e^{(s-4)\phi}  \frac{\xi(2(s-2))}{\xi(2s)} E_{\DSOIVS{0}{0}{s\mbox{-}1}{0}}\\ + \frac{2}{\xi(2s)} \sum_{\substack{p\in \mathds{Z}^8\\ (p,p)=0}} \Scal{{\rm gcd}(p)^{s-1} \sum_{n|{\rm gcd}(p) } n^{2(2-s)}}   \frac{e^{-2\phi}}{|Z(p)|} E_{\DSLIII{s\mbox{-}1}00}(v_p) K_{s{-}2}(2\pi  e^{-\phi} |Z(p)|)e^{2\pi i ( p,a)}  \end{multline}
where $v_p$ parametrizes the $SL(4)\subset SO(5,5)$ that stabilises $p$.

\end{document}